\documentclass[12pt]{iopart}

\newcommand{\bra}[1]{\langle #1 |}
\newcommand{\ket}[1]{| #1 \rangle}

\newcommand{\ua}[0]{{{\uparrow}}}
\newcommand{\da}[0]{{{\downarrow}}}
\newcommand{\md}[0]{ {\rm media}}

\newcommand{\env}[0]{ {\rm env}}
\newcommand{\tot}[0]{ {\rm tot}}
\newcommand{\eff}[0]{ {\rm eff}}
\newcommand{\cop}[0]{ {\rm couple}}

\usepackage{graphicx}
\usepackage{subfigure}
\usepackage{color}
\begin{document}

\title[  ]{General conditions for the generation of long-distance entanglement}

\author{Tomotaka Kuwahara}

\address{Department of Physics, The University of Tokyo, Komaba, Meguro, Tokyo 153-8505}
\ead{tomotaka@iis.u-tokyo.ac.jp}

\begin{abstract}

We generally investigate necessary conditions for the generation of the long-distance entanglement.
We consider a quantum system in which a system mediates the indirect interaction between two spins, which we refer to as probe spins.
First, we weaken the coupling between each probe spin and the mediator system to the infinitesimal strength in order to generate the long-distance entanglement.
We give two necessary conditions for the mediator system to generate the long-distance entanglement. 
We prove that the indirect interaction cannot generate the entanglement if it is `classical.'
We also give a necessary condition for the effective fields on the probe spins to satisfy. 
Second, we generate the long-distance entanglement by the use of only external fields.
We show that external fields on the adjacent spins to the probes are necessary in addition to external fields on the probe spins. 
Finally, we consider the cases where the coupling strength between each probe spin and the mediator system is finite.
In particular, we show two examples where the external fields on the mediator system highly enhance  the long-distance entanglement.

\end{abstract}

\maketitle

\section{Introduction} \label{Introduction}
\begin{figure}
\centering
\includegraphics[clip, scale=0.7]{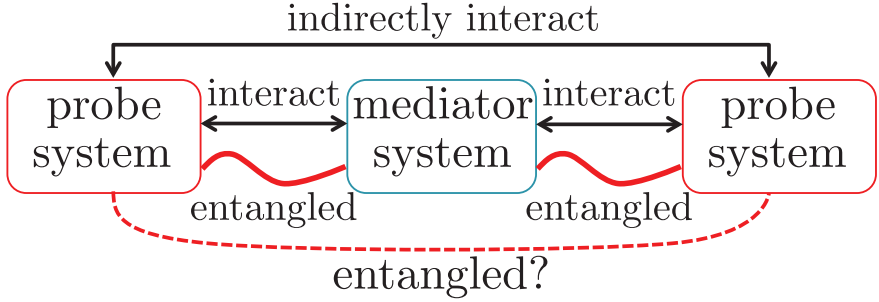}
\caption{Indirectly interacting systems. 
The two probe systems respectively interact with the mediator system. 
They can indirectly interact with each other via the mediator system though the two probe systems do not directly interact with each other.The entanglement between each probe system and the mediator system decreases the purity of the probe systems.}
\label{fig:Considering_system}
\end{figure}

The generation of the quantum entanglement~\cite{horodecki}  is of great importance because the entanglement plays an essential role in quantum information processing~\cite{Nielsen,QIBennett}.
Many people have sought efficient generation of the entanglement between remote quantum systems~\cite{Amico}.
Let us consider that two systems (probe systems) indirectly interact with each other via another quantum system (mediator system);  we summarize in Table~\ref{Table:definition} the definition of the original terms in our paper.
In this case, the entanglement of each probe system with the mediator system would decrease the purity of the probe systems (Fig.~\ref{fig:Considering_system}).
As a result, the entanglement between the two probe systems rapidly decays and vanishes as the probe systems get separated distantly. 
Indeed, we cannot usually obtain the entanglement between systems far apart even in the low-temperature limit~\cite{Arnesen,Osborne}. 
For the generation of the entanglement between the remotely separated probe systems, we therefore need to suppress the entanglement between each probe system and the mediator system.
It then appears to be a dilemma; we need to weaken the interaction between each probe system and the mediator system, and yet we need to keep the indirect interaction between the probe systems.

A recent study~\cite{LDE1}, however, reported successful generation of the entanglement between a far-separated spin pair in specific models;  such an entanglement is referred to as the \textit{long-distance entanglement}.
In their models, the entanglement with the mediator system is suppressed enough although the two spins still indirectly interact with each other via the mediator system.
After the first paper on the long-distance entanglement, various systems have turned out to be usable for generating the long-distance entanglement~\cite{LDE2,LDE3,LDE4,LDE5,LDE6,modular,LDE7,LDE8,LDE9,LDE10}.
In experiments, we will have to prepare a quantum system which we can control the system parameters.
 It is expected that we can generate the long-distance entanglement by the use of the optical lattice~\cite{Op_lattice,Op_lattice3}.
Because of its usefulness, the theoretical and the experimental studies have rapidly progressed recently~\cite{Op_lattice2}.
In order to utilize the long-distance entanglement for practical applications, however, there are still many problems to solve.
In the present paper, we find answers to the following questions on the long-distance entanglement:
\begin{enumerate}
\item{}What are the conditions for a mediator system to generate the long-distance entanglement?
\item{}What are the conditions to generate the long-distance entanglement by the use of external fields only on the probe systems?
\item{}In what ways can we enhance the capability of the mediator system to generate the long-distance entanglement?
\end{enumerate}
These questions are closely related to practical applications.

   \begin{table}[t]
\begin{center}
\begin{tabular}{ll}
\hline
 Term &  Definition \\  \hline
Probe spins & The spins between which we mainly consider the entanglement.  \\ 
 Mediator system& The quantum system which mediates  the indirect interaction  \\
 &between two probe spins.\\
Local fields& The external fields on the probe spins. \\
&The Hamiltonian is given in~(\ref{Definition_of_interaction_Hamiltonian}).\\  \hline
\end{tabular}
\caption{Definition of the original terms in the paper.}
\label{Table:definition}
\end{center}
\end{table}

One of the most popular methods of the generation of the long-distance entanglement is to weaken the coupling between each probe system and the mediator system to the infinitesimal strength~\cite{LDE1,LDE2,LDE3,LDE4,LDE5}.
It has been shown for several models that we obtain the maximum entanglement between the far-separated spins in the limit of the weak coupling.
In some models, analytical calculations of the long-distance entanglement were possible and important quantities such as the first excitation energy were obtained in Refs.~\cite{LDE2,LDE3,LDE4}.
In Ref.~\cite{LDE3}, a mathematical analysis was also given as to why we can generate the strong entanglement in this framework.
In Ref.~\cite{LDE5}, experimental realization of the long-distance entanglement has been theoretically discussed in the optical system.
However, there is no analytical argument on a general condition to generate the long-distance entanglement, which is one of the most important problems in discussing the usefulness of the long-distance entanglement.

It is also an important question whether we can generate the long-distance entanglement with external fields or not.
It is a popular problem how much entanglement can be generated by modulating system parameters~\cite{Arnesen,Bose1,Bose2,Plastina,Plastina2,Plastina3,Plastina4,Fujinaga,Kuwahara1,Kuwahara2}. 
One of the easiest parameters which we can control freely is an external fields $\{h^\xi\}_{\xi=x,y,z}$ applied on a spin of the Pauli matrices $\{\sigma^\xi\}_{\xi=x,y,z}$ in the form of the Hamiltonian $h^x \sigma^x+h^y \sigma^y+h^z \sigma^z$.
Reference~\cite{Plastina} discussed the generation of large entanglement between distantly separated two spins in an $XX$ spin chain; they showed that the local control of the external fields is an effective method of generating a large entanglement between the separated two sites in this specific spin chain.
In general systems, however, we cannot generate the long-distance entanglement if we modulate the external fields only on the two probe spins.

\begin{figure}
\centering
\subfigure[]{
\includegraphics[clip, scale=0.6]{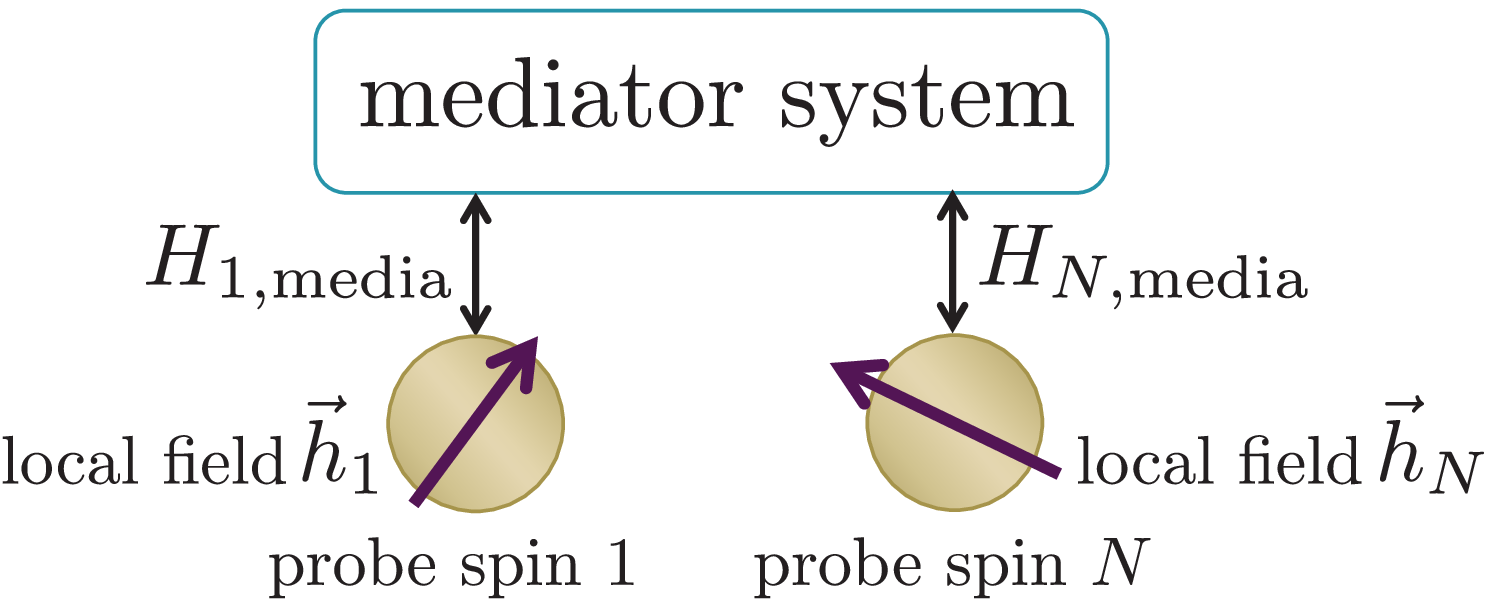}
}
\subfigure[]{
\includegraphics[clip, scale=0.6]{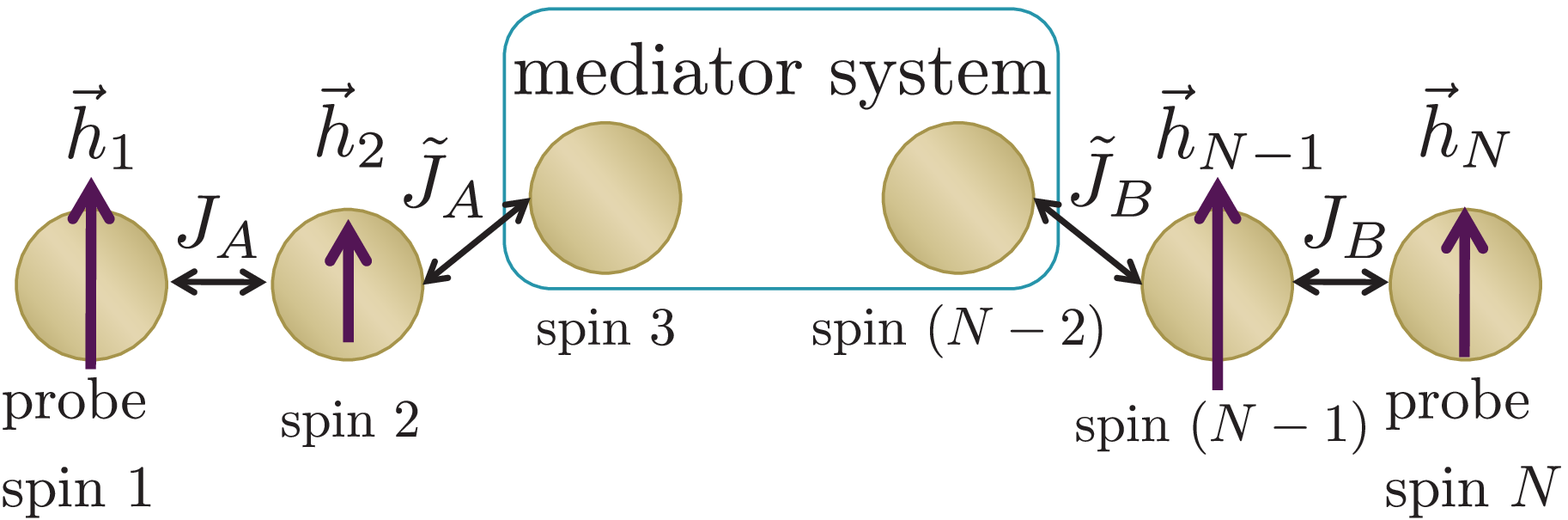}
}
\caption{(a) A schematic illustration of the generation of the long-distance entanglement in Sections~\ref{general condition for the generation of the long-distance entanglement} and \ref{the long-distance entanglement_numerical_calculation_example}. Two spins are connected to the mediator system with the coupling Hamiltonians $H_{1,\md}$ and $H_{N,\md}$.
We refer to  the spins~1 and $N$ as the probe spins and the entanglement between the probe spins as the long-distance entanglement.
In order to generate the entanglement between the probe spins, we decrease the amplitude of the coupling Hamiltonian.  We consider infinitesimally weak couplings in Section~\ref{general condition for the generation of the long-distance entanglement} and finite couplings in Section~\ref{the long-distance entanglement_numerical_calculation_example}. (b) A schematic picture of the generation of the long-distance entanglement by local fields in Section~\ref{section_the generation of the long-distance entanglement  by the local fields}. 
We connect four spins to the mediator system and consider the entanglement generation between the probe spins~1 and $N$. We apply the local fields on the four spins.
By increasing the local fields $\vec{h}_2$ and $\vec{h}_{N-1}$, we can decrease the effective coupling between each of the probe spins and the mediator system.}
\label{fig:LDE_generation}
\end{figure}

Answering the above questions, we discuss in the present paper the generation of the long-distance entanglement between the two probe spins which indirectly interact with each other through a mediator system.
Our main results are summarized as follows:
\begin{enumerate}
\item{}We first consider the system in Fig.~\ref{fig:LDE_generation} (a) and decrease the coupling strength between the mediator system and each of the probe spins~1 and $N$.
In the weak coupling limit, we give two necessary conditions for the mediator system to generate the long-distance entanglement in the forms of two sufficient conditions for the non-existence of the long-distance entanglement. First, the indirect interaction must not be `classical.'
Second, the effective fields on the probe spins must not be much larger than the indirect interaction between these two spins.
We will give mathematical expressions of these conditions below.
\item{}We next consider the generation of the long-distance entanglement with external fields. As has been expected, it is not enough to control the external fields only on the two probe spins. 
In addition to the external fields on the probe spins, we also have to control the external fields on the system adjacent to the probe spins, as is depicted in Fig.~\ref{fig:LDE_generation} (b).
We give a necessary condition for the external fields to give rise to infinitesimal effective couplings between the mediator system and each of the probe spins~1 and $N$, thus generating the long-distance entanglement between the probe spins owing to the mechanism given in the previous item (i).
\item{}We finally consider the quantum system in Fig.~\ref{fig:LDE_generation} (a) with finite couplings.
We can highly enhance the long-distance entanglement by modulating the external fields on the mediator system.
\end{enumerate}
We show these results analytically and numerically.

This paper consists of the following sections. In section~\ref{Section_General formulation of the long-distance entanglement}, we review the general framework of the generation of the long-distance entanglement.
Sections~\ref{general condition for the generation of the long-distance entanglement}, \ref{section_the generation of the long-distance entanglement  by the local fields} and \ref{the long-distance entanglement_numerical_calculation_example} describe the items (i), (ii) and (iii) above, respectively. 
In Section~\ref{conclusion}, a discussion concludes the paper.


\section{General framework of the generation of the long-distance entanglement} \label{Section_General formulation of the long-distance entanglement}

In the present section, we review the general framework of the generation of the long-distance entanglement~\cite{LDE1,LDE2,LDE3,LDE4,LDE5}.
We consider a quantum system in which two `probe' spins  are connected to an arbitrary mediator system (Fig.~\ref{fig:LDE_generation} (a)); that is, 
\begin{eqnarray}
H_{\tot}=H_{\rm int}  + H_{\rm LF}    , \label{the long-distance entanglement_Hamiltonian_total}
\end{eqnarray} 
with
\begin{eqnarray}
&H_{\rm int} =  H_\md + H_{{\rm couple}},  \nonumber \\
&H_{{\rm couple}}=\sum_{i=x,y,z}\bigl( \sigma_1^i  \otimes  H_{1,\md}^i  + \sigma_N^i  \otimes  H_{N,\md}^i \bigr) , \nonumber \\
&H_{\rm LF}= \sum_{i=x,y,z}\bigl( h_{1}^{i} \sigma_1^i   + h_{N}^{i}  \sigma_N^j \bigr),      \label{Definition_of_interaction_Hamiltonian}
\end{eqnarray} 
where ${\{\sigma^i\}_{i=x,y,z}}$ are the Pauli matrices of $S=1/2$ spins, while $\{H_{1,\md}^i,H_{N,\md}^i\}_{i=x,y,z}$ and $H_{{\rm media}}$ are arbitrary (but not the identity) $2^{N-2}$-dimensional Hamiltonians which do not include the spins~1 and $N$.
We assume that the ground state of $H_{{\rm media}}$ is not degenerate.
We define $H_{\rm int}$ as $H_{\tot}-H_{\rm LF}$, which characterizes the indirect interaction between the spins~1 and $N$.
Note that the external fields on the \textit{mediator system} are also included in the Hamiltonian $H_{\rm int}$; it means that the indirect interaction between the spins~1 and $N$ depends on the external fields on the mediator system. 
We hereafter refer to the spins~1 and $N$ as the probe spins and to the external fields $\{h_{1}^{i},h_{N}^{i}\}_{i=x,y,z}$ as the local fields (Table~\ref{Table:definition}).
We also refer to the entanglement between the probe spins~1 and $N$ as the long-distance entanglement.
In fact, the long-distance entanglement is usually defined as the entanglement between infinitely separated two spins~\cite{LDE1}; if the distance between the two probe spins is large but finite, the entanglement between such a spin pair is often referred to as the quasi-long-distance entanglement~\cite{LDE2}.  In the present paper, however, we refer to all the entanglement which is generated by the indirect interaction as the long-distance entanglement for simplicity.

Let us consider the problem of enhancing the ground-state entanglement between the spins 1 and $N$ by modulating the coupling Hamiltonian $H_{{\rm couple}}$. 
The density matrix of the total system in the ground state is given by 
\begin{eqnarray}
\rho_{\tot} &=\lim_{\beta \to \infty} \frac{e^{-\beta H_{\tot}}}{Z_{\tot}(\beta)},     \label{density matrix}
\end{eqnarray}
where $\beta$ is the inverse temperature $(k_{\rm B}T)^{-1}$ with $k_{\rm B}$ the Boltzmann constant and  ${Z_{\tot}(\beta) ={\rm tr}( e^{-\beta H_{\tot}})}$ is the partition function.
The density matrix of the probe spins $1$ and $N$ is
\begin{eqnarray}
\rho_{1N} &= \tr_{1N}\rho_{\tot},  \label{density_matrix_spin1_spinN}
\end{eqnarray}
where $\tr_{1N}$ denotes the trace operation on the system \textit{except} the probe  spins~$1$ and $N$.
In order to quantify the entanglement, we here adopt the concurrence~\cite{Wootter}, which is most commonly used as an entanglement measure.
The concurrence $C(\rho_{1N})$ is defined as follows:
\begin{eqnarray} 
C(\rho_{1N}) \equiv \max(\lambda_1-\lambda_2-\lambda_3-\lambda_4,0)  ,     \label{definition_of_the_concurrence}
\end{eqnarray} 
where $\{\lambda_n\}_{n=1}^4$ are the eigenvalues of the $4\times4$ matrix
\begin{eqnarray} 
\sqrt{\rho_{1N} (\sigma_1^y \otimes\sigma_N^y) \rho_{1N}^{\ast}(\sigma_1^y\otimes \sigma_N^y) }
\end{eqnarray} 
in the non-ascending order $\lambda_1\ge\lambda_2\ge\lambda_3\ge\lambda_4$.
Note that $C(\rho_{1N})>0$ is a necessary and sufficient condition for the existence of the entanglement.

A popular method of generating the long-distance entanglement in the ground state is to reduce the coupling Hamiltonian $H_{{\rm couple}}$ to zero~\cite{LDE1,LDE2,LDE3,LDE4,LDE5}; namely
\begin{eqnarray}
\|H_{{\rm couple}} \| \rightarrow 0   ,  \label{the long-distance entanglement_generation}
\end{eqnarray} 
where $\|\cdots\|$ denotes the matrix norm.
It has been shown that the long-distance entanglement is generated in the limit (\ref{the long-distance entanglement_generation}) as
\begin{eqnarray} 
\lim_{\|H_{{\rm couple}} \| \to0} C(\rho_{1N}) > 0  \label{condition_for_the long-distance entanglement_generated}
\end{eqnarray} 
in some quantum systems~\cite{{LDE1,LDE2,LDE3,LDE4,LDE5}}, for example, an \textit{XX} spin chain.

  \begin{figure}
\centering
\subfigure[Non-zero entanglement]{
\includegraphics[clip, scale=0.6]{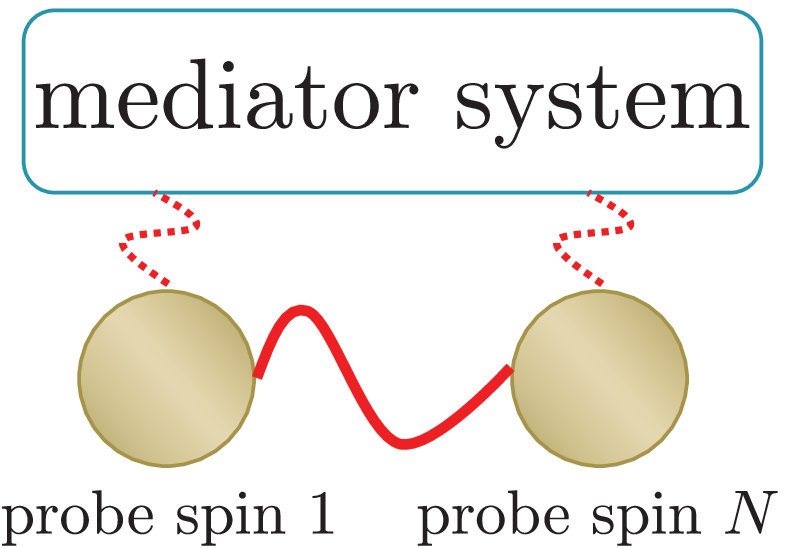}
}
\subfigure[No entanglement]{
\includegraphics[clip, scale=0.6]{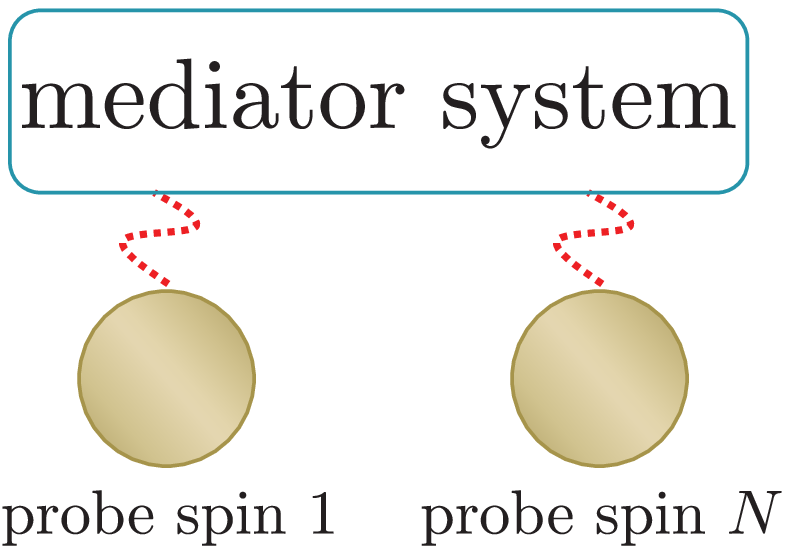}
}
\caption{Comparison of the entangled ground state and the non-entangled ground state. In the limit of weak coupling, the probe spins~1 and $N$ cannot entangle strongly with the mediator system because of the monogamy.
Therefore, we can consider the two cases (a) and (b).
In the case (a), the energy of the entangled state is less than the non-entangled state, while in the case (b), the energy of the non-entangled state is less than the entangled state. 
}
\label{fig:Explanation_LDE}
\end{figure}

Let us explain qualitatively why we can generate the long-distance entanglement in the limit~(\ref{the long-distance entanglement_generation}).
A key feature for the generation of the long-distance entanglement is the monogamy~\cite{monogamy}, which is a property that multiple pairs cannot share a strong entanglement simultaneously.
Because of the monogamy, if the components of the mediator system are entangled strongly with each other, a large entanglement cannot be generated between each probe spin and the mediator system.
In the limit~(\ref{the long-distance entanglement_generation}) of weak coupling, the components of the mediator system may be entangled strongly with each other, but are scarcely entangled with each probe spin.
Then, the probe spins can be entangled strongly only with each other.

It then appears to be claimed that the probe spins are entangled without the interaction, which is obviously not the case.
The fact is that the interaction is weak but not zero, and hence the quantum correlation between the two probe spins can decrease the energy of the total system.
Then the ground state may support the entanglement  between the probe spins (Fig.~\ref{fig:Explanation_LDE}).
We can describe the indirect interaction in terms of the effective Hamiltonian in the limit~(\ref{the long-distance entanglement_generation}). We will give the effective Hamiltonian in Section~\ref{Section_Effective_field} in order to obtain general properties of the long-distance entanglement.

The next question is how to realize the weak-coupling limit~(\ref{the long-distance entanglement_generation}) in realistic situations.
We answer it by modulating local fields in the system in Fig.~\ref{fig:LDE_generation} (b), extending Fig.~\ref{fig:LDE_generation} (a) slightly.
By increasing the local fields on the spins~2 and $N-1$, we can effectively decrease the coupling between the mediator system and each of the probe spins~1 and $N$.
In fact, as will be shown in  Section~\ref{section_the generation of the long-distance entanglement  by the local fields}, we cannot achieve the long-distance entanglement by simply letting $|\vec{h}_2|$ and $|\vec{h}_{N-1}|$ to infinity but also by adjusting $\vec{h}_1$ and $\vec{h}_N$ accordingly. 
The merit of using the external fields is that we can control them more easily and precisely than the coupling interactions.
We then consider in Section~4 the following Hamiltonian: 
\begin{eqnarray}
H_{\tot}=& \sum_{i,j=x,y,z}\bigl( J_{A}^{i,j} \sigma_1^i  \sigma_2^j  + \tilde{J}_{A}^{i,j} \sigma_2^i  \sigma_3^j   +\tilde{J}_{B}^{i,j} \sigma_{N-2}^i  \sigma_{N-1}^j  +J_{B}^{i,j} \sigma_{N-1}^i \sigma_N^j \bigr)\nonumber \\
&+\sum_{i=x,y,z} \bigl( h_1^{i} \sigma_1^i  +  h_2^{i} \sigma_2^i  +  h_{N-1}^{i} \sigma_{N-1}^i  +  h_N^{i} \sigma_N^i  \bigr)  +H_{{\rm media}}  ,   \label{the long-distance entanglement_Hamiltonian_maximum_field}
\end{eqnarray} 
where the Hamiltonian $H_{{\rm media}}$ is now an arbitrary $2^{N-4}$-dimensional Hamiltonian and we assume that the ground state of $H_{{\rm media}}$ is not degenerate.
We modulate only the local fields $\{h_1^{i},h_2^{i},h_{N-1}^{i},h_N^{i}\}_{i=x,y,z}$ in order to generate the entanglement between the spins 1 and $N$. 
In Section~\ref{section_the generation of the long-distance entanglement  by the local fields}, we will clarify the exact correspondence between the decrease of the interaction and the increase of the local fields.


\section{General conditions for the generation of the long-distance entanglement} \label{general condition for the generation of the long-distance entanglement}

In the previous section, we presented general statements on the generation of the long-distance entanglement.
However, they do not tell us whether the mediator system contributes to the generation of the long-distance entanglement or not.
We here present two cases in which we cannot generate the long-distance entanglement.
In the first case, indirect interactions in a particular class cannot generate the entanglement at all; we refer to such an interaction as the `classical' interaction.
In the second case, the effective fields on the probe spins seriously destroy the entanglement.
These constitute sufficient conditions for the non-existence of the long-distance entanglement, thereby giving two necessary conditions for its existence as their contrapositions. 
Throughout the present section, we consider the system~(\ref{the long-distance entanglement_Hamiltonian_total}) in Fig.~\ref{fig:LDE_generation} (a).


\subsection{Classical interaction}
The first necessary condition for the generation of the long-distance entanglement  is that the indirect interaction between the spins~1 and $N$ must not be a `classical' interaction.
We define that the indirect interaction between the probe spins~1 and $N$ is `classical'  if there exists the following separation of the Hamiltonian $H_{{\rm int}}$ in~(\ref{the long-distance entanglement_Hamiltonian_total}):
\begin{eqnarray}
&H_{{\rm int}} = H_{A} (\sigma_1) + H_{B} (\sigma_N)  \nonumber \\
&{\rm with}  \quad [  H_{A} (\sigma_1) , H_{B} (\sigma_N)  ] =0. \label{Sep_cond}
\end{eqnarray}
As has been shown in Eq.~(\ref{the long-distance entanglement_Hamiltonian_total}), the Hamiltonian $H_{{\rm int}}$ is defined as the Hamiltonian except the external fields on the spins~$1$ and $N$.
Note that $H_A$ does not contain $\sigma_N$ nor $H_B$ contains $\sigma_1$. 
We can prove the following theorem. 

 \textit{Theorem~1}:  
If the interaction Hamiltonian $H_{{\rm int}}$ is `classical,' we cannot generate the entanglement between the spins~$\sigma_1$ and $\sigma_N$ for any values of the local fields $\{h_1^i\}_{i=x,y,z}$ and $\{h_N^i\}_{i=x,y,z}$;  namely
\begin{eqnarray} 
C(\rho_{1N})=0  ,
\end{eqnarray} 
where the density matrix $\rho_{1N}$ is defined in Eqs.~(\ref{density matrix}) and (\ref{density_matrix_spin1_spinN}) and the concurrence $C(\rho_{1N})$ is defined in Eq.~(\ref{definition_of_the_concurrence}).

\textit{Comments}:
For example, we can separate the following Hamiltonian in the form (\ref{Sep_cond}):
\begin{eqnarray}
H_{{\rm int}}= J_1^z \sigma_1^z   \sigma_{2}^z+  J_2^x \sigma_2^x   \sigma_{3}^x +J_3^x \sigma_3^x   \sigma_{4}^x , \label{classical_interaction_one_example_zero}
\end{eqnarray}
where the spin pair $(1,2)$ interact with each other through the Ising interaction along the $z$-axis, while the spin pairs $(2,3)$ and $(3,4)$ interact with each other through the Ising interaction along the $x$-axis.
We can separate this Hamiltonian into $ H_{A} (\sigma_1)$ and $H_{B} (\sigma_4)$ as
\begin{eqnarray}
H_{A} (\sigma_1) &=J_1^z \sigma_1^z   \sigma_{2}^z+  J_2^x \sigma_2^x   \sigma_{3}^x,  \nonumber \\
H_{B} (\sigma_4)  &= J_3^x \sigma_3^x   \sigma_{4}^x   . \label{classical_interaction_one_example}
\end{eqnarray}
These Hamiltonians satisfy the condition $[H_{A} (\sigma_1) , H_{B} (\sigma_4)] = 0$, and hence we cannot generate the entanglement between the probe spins~$1$ and $4$ in this system for any values of the local fields $\{h_1^i\}_{i=x,y,z}$ and $\{h_4^i\}_{i=x,y,z}$.
Note that the spins~1 and 4 are classically correlated with each other.
If we replace $J_3^x \sigma_3^x   \sigma_{4}^x $ in Eq.~(\ref{classical_interaction_one_example_zero}) by $J_3^z \sigma_3^z   \sigma_{4}^z $ as
\begin{eqnarray}
H_{{\rm int}}=J_1^z \sigma_1^z   \sigma_{2}^z+  J_2^x \sigma_2^x   \sigma_{3}^x+J_3^z \sigma_3^z   \sigma_{4}^z ,\label{classical_to_Nonclassical_interaction_one_example}
\end{eqnarray} 
we cannot separate the Hamiltonian $H_{{\rm int}}$ into the forms of $ H_{A} (\sigma_1)$ and $H_{B} (\sigma_4)$ which satisfy $[H_{A} (\sigma_1) , H_{B} (\sigma_4)] = 0$ anymore, and hence the spins~1 and 4 can entangle with each other.

It is worth noting that by the external fields on the mediator system the interaction Hamiltonian~$H_{\rm int}$ can be transformed from a `classical' one to a `non-classical' one, that is, the entanglement generation becomes possible.
For example, if we add the external field $h_3^z \sigma_3^z$ on the spin~3 in Eq.~(\ref{classical_interaction_one_example_zero}), we cannot separate the Hamiltonian as in Eq.~(\ref{classical_interaction_one_example}).
We will also show in the section~5 that the external fields on the mediator system can enhance the capability of the interaction Hamiltonian~$H_{\rm int}$.

Eigenstates can have the entanglement even if the condition~(\ref{Sep_cond}) is satisfied.
For example, the Hamiltonian for the probe spins~1 and 3,
\begin{eqnarray}
&H_{\tot}=H_{{\rm int}} + h_1^z\sigma_1^z+ h_3^z\sigma_3^z, \nonumber \\
&H_{{\rm int}}=J_1^z \sigma_1^z   \sigma_{2}^z+ J_2^z \sigma_2^z   \sigma_{3}^z  \label{degenerate_classical_interaction_Hamiltonian}
\end{eqnarray}
with $h_1^z=h_3^z=J_1^z=J_2^z$ satisfies the condition (\ref{Sep_cond}), but it has an eigenstate $(\ket{{\ua_1\ua_2\ua_3}}+\ket{{\da_1\ua_2\da_3}})/\sqrt{2}$, which is highly entangled.
Mixing of all the eigenstates with the Boltzmann weight always destroys the entanglement between the probe spins.

Finally, under this condition, for appropriate values of the local fields we can generate the quantum discord, which is one of the non-classical correlations. We discuss the quantum discord in \ref{appendixA_chapter3}.


\textit{Proof}: 
We prove the following equality under the condition~(\ref{Sep_cond}):
\begin{eqnarray}
 \tr_{1N}( e^{-\beta H_{\tot}})=   \sum_{\tilde{n}} \rho_1^{\tilde{n}} \otimes  \rho_N^{\tilde{n}}, \label{mixture_of_the_product_state}
\end{eqnarray}
where $\tr_{1N}$ denotes the trace operation on the system except the probe spins~1 and $N$, and the density matrices $\rho_1^{\tilde{n}}$ and $\rho_N^{\tilde{n}}$ are physical states, namely, positive matrices.
Then, the spins~1 and $N$ are not entangled with each other by definition.
By proving Eq.~(\ref{mixture_of_the_product_state}), we can also prove in the limit $\beta\to\infty$ that the density matrix~(\ref{density_matrix_spin1_spinN}) is decomposed into the mixture of the product states.

First, under the condition (\ref{Sep_cond}), we can decompose the density matrix as follows:
\begin{eqnarray}
 e^{-\beta H_{\tot}} = e^{-\beta H_{A} (\sigma_1)} e^{-\beta H_{B} (\sigma_N)}.
\end{eqnarray}
We can express $ e^{-\beta H_{A} (\sigma_1)}$ and $e^{-\beta H_{B} (\sigma_N)}$ as
\begin{eqnarray} 
&e^{-\beta H_{A} (\sigma_1)} = \sum_{\mu=0,x,y,z}  \sigma_1^\mu \otimes \rho_{{\rm media}}^{1\mu} \otimes I_N ,\nonumber \\
&e^{-\beta H_{B} (\sigma_N)} = \sum_{\nu=0,x,y,z}  I_1 \otimes \rho_{{\rm media}}^{N\nu} \otimes \sigma_N^\nu ,   \label{Equations_3_15}
\end{eqnarray}
where $I_1$ and $I_N$ are the identity matrices in the spaces of the spins~1 and $N$, respectively, and we define $\sigma_1^0=I_1$ and $\sigma_N^0=I_N$.
We also define that the matrices $\rho_{{\rm media}}^{1\mu}$ and $\rho_{{\rm media}}^{N\nu}$ are Hermitian operators in the mediator space.
Because $H_{A} (\sigma_1)$ and $H_{B} (\sigma_N)$ are assumed to commute with each other, the matrices $ e^{-\beta H_{A} (\sigma_1)}$ and $e^{-\beta H_{B} (\sigma_N)}$  also commute with each other.
Therefore, we obtain the following equation:
\begin{eqnarray} 
{\rm tr}^{1N}\bigl[ \sigma_1^\mu \otimes \sigma_N^\nu e^{-\beta H_{A} (\sigma_1)} e^{-\beta H_{B} (\sigma_N)} \bigr]  &= {\rm tr}^{1N}\bigl[\sigma_1^\mu \otimes \sigma_N^\nu e^{-\beta H_{B} (\sigma_N)} e^{-\beta H_{A} (\sigma_1)} \bigr],
\end{eqnarray}
where $\tr^{1N}$ denotes the trace operation only on the spins~1 and $N$. From this equation we obtain
\begin{eqnarray}
\rho_{{\rm media}}^{1\mu}  \rho_{{\rm media}}^{N\nu} = \rho_{{\rm media}}^{N\nu}  \rho_{{\rm media}}^{1\mu} \label{exchange_rho1_rhoN}
\end{eqnarray}
for $\mu,\nu=0,x,y,z$.
Therefore, the matrices $\rho_{{\rm media}}^{1\mu}$ and $\rho_{{\rm media}}^{N\nu}$ have simultaneous eigenstates.
Then, we can express $\rho_{{\rm media}}^{1\mu}$ and $\rho_{{\rm media}}^{N\nu}$ as
\begin{eqnarray}
&\rho_{{\rm media}}^{1\mu} =\sum_{n=1}^{2^{N-2}} \lambda_{\mu}^n \ket{n,\mu_1,\nu_N} \bra{n,\mu_1,\nu_N}                \label{rho_media1}
\end{eqnarray}
and
\begin{eqnarray}
&\rho_{{\rm media}}^{N\nu} = \sum_{n=1}^{2^{N-2}} \tau_{\nu}^n \ket{n,\mu_1,\nu_N} \bra{n,\mu_1,\nu_N}   ,  \label{rho_mediaN}
\end{eqnarray}
where $\{\ket{n,\mu_1,\nu_N}\}$ are $2^{N-2}$ pieces of the simultaneous eigenstates of $\rho_{{\rm media}}^{1\mu}$ and $\rho_{{\rm media}}^{N\nu}$. 
As a result, we obtain
\begin{eqnarray}
\fl&e^{-\beta H_{A} (\sigma_1)} e^{-\beta H_{B} (\sigma_N)} \nonumber \\
\fl=&  \biggl (\sum_{n,\mu} \lambda_{\mu}^n \sigma_1^\mu \otimes \ket{n,\mu_1,\nu'_N} \bra{n,\mu_1,\nu'_N}  \otimes I_N \biggr) \biggl (\sum_{n',\nu} \tau_{\nu}^n I_1 \otimes \ket{n',\mu'_1,\nu_N} \bra{n',\mu'_1,\nu_N}  \otimes \sigma_N^{\nu} \biggr)   \nonumber \\
\fl=&\sum_{n,\mu,\nu} \lambda_{\mu}^n \tau_{\nu}^n \sigma_1^\mu \otimes \ket{n,\mu_1,\nu_N} \bra{n,\mu_1,\nu_N}  \otimes \sigma_N^\nu ,
\end{eqnarray}
where the indices $\mu'$ and $\nu'$  in the first line can be arbitrarily chosen $(\nu',\mu'=0,x,y,z)$, and hence we choose $\mu'$ and $\nu'$ in accordance with $\mu$ and $\nu$.
By tracing out the mediator space, we have
\begin{eqnarray}
{\rm tr}_{1N}e^{-\beta H_{A} (\sigma_1)} e^{-\beta H_{B} (\sigma_N)} &=  \sum_{n,\mu,\nu} \lambda_{\mu}^n \tau_{\nu}^n \sigma_1^\mu \otimes \sigma_N^\nu \nonumber \\
&=\sum_{n} \Bigl(\sum_\mu  \lambda_{\mu}^n \sigma_1^\mu \Bigr) \otimes  \Bigl( \sum_\nu \tau_{\nu}^n \sigma_N^\nu \Bigr).    \label{theorem1_decompose_1}
\end{eqnarray}
At this moment, we cannot say that $\sum_\mu  \lambda_{\mu}^n \sigma_1^\mu$ and $\sum_\tau \nu_{\nu}^n \sigma_N^\nu$ are necessarily physical states, namely, positive matrices.
In the following, we prove that Eq.~(\ref{theorem1_decompose_1}) can be reduced to the mixture of the product states as in the form (\ref{mixture_of_the_product_state}).

For the purpose, we should pay attention to the degeneracies of the matrices $\rho_\mathrm{media}^{1\mu}$ and $\rho_\mathrm{media}^{N\nu}$.
In fact, if there are no degeneracies in the eigenspaces of all these matrices for $\mu,\nu=0,x,y,z$, we can easily prove that each of $\sum_\mu\lambda_\mu^n\sigma_1^\mu$ and $\sum_\nu\tau_\nu^n\sigma_N^\nu$ ($n=1,2,\ldots,2^{N-2}$) in Eq.~(\ref{theorem1_decompose_1}) is a positive matrix.
Since the matrices $\rho_\mathrm{media}^{1\mu}$ and $\rho_\mathrm{media}^{N\nu}$ commute with each other as well as $\rho_\mathrm{media}^{1\mu'}$ and $\rho_\mathrm{media}^{N\nu}$ do,  the matrices $\rho_\mathrm{media}^{1\mu}$ and $\rho_\mathrm{media}^{1\mu'}$ should also have simultaneous eigenstates if there are no degeneracies.
If there are absolutely no degeneracies in all eigenspaces of $\rho_\mathrm{media}^{1\mu}$ and $\rho_\mathrm{media}^{N\nu}$ ($\mu,\nu=0,x,y,z$), we have an orthonormal set of $2^{N-2}$ pieces of states $|n\rangle$, each of which is the simultaneous eigenstate $|n,\mu_1,\nu_N\rangle$ for all of $\mu,\nu=0,x,y,z$.
Then, we have from Eq.~(\ref{Equations_3_15})
\begin{eqnarray}
\sum_{\mu=0,x,y,z}\lambda_\mu^n\sigma_1^\mu&=\mathrm{tr}_{1N}(e^{-\beta H_A(\sigma_1)}|n\rangle\langle n|),
\\
\sum_{\nu=0,x,y,z}\tau_\nu^n\sigma_N^\nu&=\mathrm{tr}_{1N}(e^{-\beta H_B(\sigma_N)}|n\rangle\langle n|),
\end{eqnarray}
for $n=1,2,\ldots, 2^{N-2}$.
This means that each of $\sum_\mu\lambda_\mu^n\sigma_1^\mu$ and $\sum_\nu\tau_\nu^n\sigma_N^\nu$ ($n=1,2,\ldots,2^{N-2}$) is a positive matrix, and hence Eq.~(\ref{theorem1_decompose_1}) indeed takes the form~(\ref{mixture_of_the_product_state}).

\begin{figure}
\centering
\includegraphics[clip, scale=0.6]{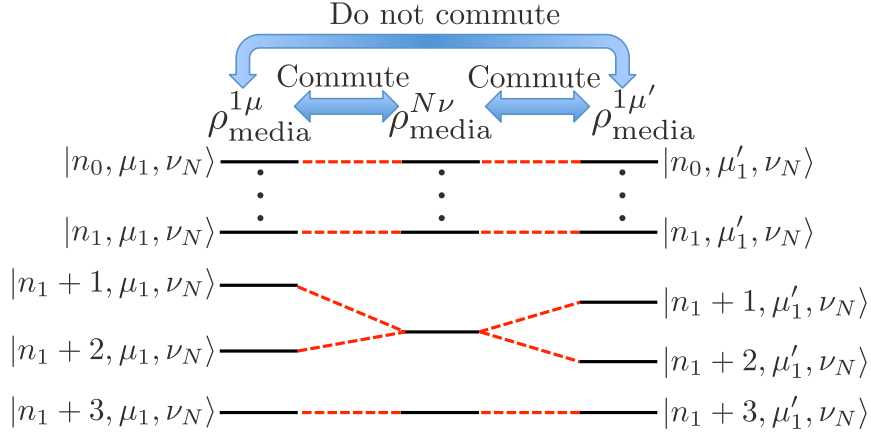}
\caption{A schematic picture of the eigenstates of $\rho_\mathrm{media}^{1\mu}$, $\rho_\mathrm{media}^{N\nu}$ and $\rho_\mathrm{media}^{1\mu'}$. 
Suppose that the $n_0$th and the $n_1$th eigenvalues of $\rho_\mathrm{media}^{N\nu}$ are not degenerate ($\tau_\nu^{n_0}\neq \tau_\nu^{n_1}$); then 
the eigenstates $\ket{n_0,\mu_1,\nu_N}$ and $\ket{n_1,\mu_1',\nu_N}$ are orthogonal to each other.
Suppose that all the eigenstates of $\rho_\mathrm{media}^{N\nu}$ are degenerate at the $(n_1+1)$th and the $(n_1+2)$th levels, namely, $\tau_\nu^{n_1+1} = \tau_\nu^{n_1+2}$ for $\nu=0,x,y,z$.
Then, any superposition of the states $\ket{n_1+1,\mu_1,\nu_N}$ and $\ket{n_1+2,\mu_1,\nu_N}$ can be the $(n_1+1)$th eigenstates of $\rho_\mathrm{media}^{N\nu}$, and hence it is possible that $\rho_\mathrm{media}^{1\mu}$ and $\rho_\mathrm{media}^{1\mu'}$ do not have the simultaneous eigenstates in the space $\{\ket{n_1+1,\mu_1,\nu_N},\ket{n_1+2,\mu_1,\nu_N}\}$. 
We then form a block which is composed of the states $\ket{n_1+1,\mu_1,\nu_N}$ and $\ket{n_1+2,\mu_1,\nu_N}$.}
\label{fig:Energy_degeneracy}
\end{figure}

If there are degeneracies in some of the eigenspaces of the matrices $\rho_\mathrm{media}^{1\mu}$ and $\rho_\mathrm{media}^{N\nu}$, there is a possibility that we cannot choose a common state $|n\rangle$ that represents the simultaneous eigenstate $|n,\mu_1,\nu_N\rangle$ for $\mu,\nu=0,x,y,z$ (Fig.~\ref{fig:Energy_degeneracy}).
Let us then inspect the degeneracies in more detail.

Suppose that the matrices $\rho_\mathrm{media}^{1\mu}$ and $\rho_\mathrm{media}^{N\nu}$ share the eigenstate $|n_0,\mu_1,\nu_N\rangle$ with the respective eigenvalues $\lambda_\mu^{n_0}$ and $\tau_\nu^{n_0}$.
We can choose the state $|n_0,\mu_1,\nu_N\rangle$ even when each of the eigenvalues $\lambda_\mu^{n_0}$ and $\tau_\nu^{n_0}$ has degeneracies in its own eigenspace.
Suppose also that the matrices $\rho_\mathrm{media}^{1\mu'}$ and $\rho_\mathrm{media}^{N\nu}$ share the eigenstate $|n_1,\mu'_1,\nu_N\rangle$ with the respective eigenvalues $\lambda_{\mu'}^{n_1}$ and $\tau_\nu^{n_1}$.
After close inspection, we can state the following:
if the eigenvalues $\tau_\nu^{n_0}$ and $\tau_\nu^{n_1}$ are not degenerate, the states $|n_0,\mu_1,\nu_N\rangle$ and $|n_1,\mu'_1,\nu_N\rangle$ are orthogonal to each other.
The only possibility that we cannot choose a common state $|n\rangle$ then occurs when all the matrices $\rho_\mathrm{media}^{1\mu}$ ($\mu=0,x,y,z$) have degeneracies in the corresponding eigenspaces and/or all the matrices $\rho_\mathrm{media}^{N\nu}$ ($\nu=0,x,y,z$) have degeneracies in the corresponding eigenspaces.

We can thereby break down the whole eigenspace into blocks.
We form a block of eigenspace in which all the matrices $\rho_\mathrm{media}^{1\mu}$ ($\mu=0,x,y,z$) have degeneracies (Case~A) and/or all the matrices $\rho_\mathrm{media}^{N\nu}$ ($\nu=0,x,y,z$) have degeneracies (Case~B).
Let us denote each block as $\mathcal{H}_{\tilde{n}}$ with the dimensionality $D_{\tilde{n}}$.
Let us choose an arbitrary orthonormal set of states $|n\rangle_{\tilde{n}}$ ($n=1,2,\ldots,D_{\tilde{n}}$) in the block $\mathcal{H}_{\tilde{n}}$.
Then we sum the terms $\sum_\mu\lambda_\mu^n\sigma_1^\mu\otimes\sum_\nu\tau_\nu^n\sigma_N^\nu$ inside each block $H_{\tilde{n}}$ to have
\begin{eqnarray}
\fl&\sum_{n: |n\rangle_{\tilde{n}} \in\mathcal{H}_{\tilde{n}}}
\sum_{\mu=0,x,y,z}\lambda_\mu^n\sigma_1^\mu \otimes \sum_{\nu=0,x,y,z}\tau_\nu^n\sigma_N^\nu \nonumber \\
\fl=&
\cases{  \tr_{1N}\bigl(e^{-\beta H_A(\sigma_1)}  |1\rangle_{\tilde{n}} \langle 1|_{\tilde{n}}\bigr)   \otimes \tr_{1N}\biggl(e^{-\beta H_B(\sigma_N)}\sum_{n=1}^{D_{\tilde{n}}}  |n\rangle_{\tilde{n}} \langle n|_{\tilde{n}}\biggr)  \quad   {\rm in \ Case~A},  \\
\tr_{1N}\biggl(e^{-\beta H_A(\sigma_1)} \sum_{n=1}^{D_{\tilde{n}}}  |n\rangle_{\tilde{n}} \langle n|_{\tilde{n}}\biggr)\otimes \tr_{1N}\bigl(e^{-\beta H_B(\sigma_N)}   |1\rangle_{\tilde{n}} \langle 1|_{\tilde{n}}\bigr)    \quad    {\rm in \ Case~B}.}  
\end{eqnarray}
This shows that Eq.~(\ref{theorem1_decompose_1}) can be summarized into the form Eq.~(\ref{mixture_of_the_product_state}), where the summation in the right-hand side of Eq.~(\ref{mixture_of_the_product_state}) is taken over the blocks $\tilde{n}$.
 Thus, Theorem~1 is proved. $\opensquare$

The contraposition of Theorem~1 gives us a necessary condition to generate the entanglement between the probe spins by the interaction~(\ref{Definition_of_interaction_Hamiltonian}). 
We can see from the proof that we can extend this theorem to be applicable to any bipartite systems which indirectly interact with each other at arbitrary temperatures.
Let us consider the interaction between the bipartite system with $\mathcal{L}$ levels; namely
\begin{eqnarray}
H_{{\rm int}}= \sum_{i=1}^{\mathcal{L}^2-1}\bigl( \tilde{\sigma}_1^i  \otimes  H_{1,\md}^i  + \tilde{\sigma}_N^i  \otimes  H_{N,\md}^i \bigr) +  H_\md ,  
\end{eqnarray} 
where $\{\tilde{\sigma}_1^i,\tilde{\sigma}_N^i\}_{i=1}^{\mathcal{L}^2-1}$ are the bases of the bipartite systems and we define $\{\tilde{\sigma}_1^0,\tilde{\sigma}_N^0\}$ as the identity matrices.
In Eq.~(\ref{Equations_3_15}), we had $\mu,\nu=0,x,y,z$ for the two $S=1/2$ spins, while we have $\mu,\nu=0,1,2,\ldots,\mathcal{L}^2-1$ for this bipartite system.
The extension of the proof is straightforward.


\subsection{Effective fields} \label{Section_Effective_field}
We next show that the effective fields on the probe spins have serious effects on the generation of the long-distance entanglement.
In order to make the discussion clear, we first give the formal expression of the effective Hamiltonian of the spins~1 and $N$.

We investigate the ground state of the total Hamiltonian $H_\tot$ of the system in Fig.~\ref{fig:LDE_generation} (a).
For this purpose, we break down the Hamiltonian~(\ref{the long-distance entanglement_Hamiltonian_total}) as follows:
\begin{eqnarray}
&H_{\tot}=H_0 +H_1 ,\quad  H_0\equiv H_{{\rm media}},  \nonumber \\
&H_1\equiv  H_{{\rm couple}}+  H_{{\rm LF}}       . \label{Separation_Hamiltonian_effective_field}
\end{eqnarray} 
We define the ground state of $H_0$ as $\ket{{\psi_0^{\md}}}$ with the eigenvalue $E_0^\md$.
Because we assumed that the ground state of the mediator system is not degenerate, we also define the first excitation energy of $H_{{\rm media}}$ as $\delta E_1^{\md}$ ($>0$).

We then assume
\begin{eqnarray}
\|H_1\| \ll \delta E_1^{\md} \label{Condition_enough_small}
\end{eqnarray}
and consider the term $H_1$ as perturbation.
The unperturbed ground state is given by
\begin{eqnarray}
P_0 \equiv I_{1N} \otimes \ket{{\psi_0^\md}}\bra{{\psi_0^\md}},
\end{eqnarray}
where $I_{1N}$ is the identity matrix of the spins~1 and $N$.

In order to define the effective Hamiltonian, we consider a Green's function
\begin{eqnarray}
\frac{1}{E-P_0 H_\tot P_0  } = P_0 \frac{1}{E-H_\tot} P_0  \label{Green_function_effective}
\end{eqnarray}
and define the effective Hamiltonian as
\begin{eqnarray}
\frac{1}{E-P_0 H_\tot P_0  } \equiv \frac{1}{E-H^\eff}.  \label{Green_function_effective122}
\end{eqnarray}
We calculate the effective Hamiltonian by expanding Eq.~(\ref{Green_function_effective}) with respect to $H_1$ to  have
\begin{eqnarray} 
P_0 \frac{1}{E-H_\tot} P_0 &=P_0 \frac{1}{E-H_0 - H_1} P_0 \nonumber \\
&=P_0 \frac{1}{E-H_0} P_0  +   P_0  \frac{1}{E-H_0}  H_1 \frac{1}{E-H_0}  P_0  \nonumber \\
&\quad +P_0  \frac{1}{E-H_0}H_1 \frac{1}{E-H_0} H_1 \frac{1}{E-H_0}    P_0 + O\bigl(\| H_1\|^3 \bigr) . \label{Green_function_effective1}
\end{eqnarray}
We can rewrite Eq.~(\ref{Green_function_effective1}) as
\begin{eqnarray} 
\fl P_0 \frac{1}{E-H_\tot} P_0&= \frac{1}{E-P_0H_0P_0 -P_0 H_1P_0 - P_0  H_1 Q_0 \frac{1}{E-H_0}Q_0 H_1 P_0 } + O\bigl(\| H_1\|^3 \bigr) , \label{Green_function_effective2}
\end{eqnarray}
where $Q_0\equiv I_\tot - P_0$ with $I_\tot$ the identity matrix of the total system.
We can indeed confirm Eq.~(\ref{Green_function_effective2}) by expanding it.
The pole of the Green's function (\ref{Green_function_effective2}) is given by $E=E_0^\md + O(\|H_1\|)$, and hence we substitute $E=E_0^\md$ into the term  $ P_0  H_1 Q_0 \frac{1}{E-H_0}Q_0 H_1 P_0 $.
We thereby obtain the effective Hamiltonian around the ground state $E_0^{\md}$ as
\begin{eqnarray} 
H^\eff =P_0 H_0 P_0+P_0 H_1P_0 + P_0 H_1 Q_0 \frac{1}{E_0^\md-H_0}Q_0  H_1 P_0  \label{Effective_Hamiltonian_green}
\end{eqnarray}

Let us calculate the effective Hamiltonian~(\ref{Effective_Hamiltonian_green}) using (\ref{Separation_Hamiltonian_effective_field}).
Because the Hamiltonian $H_{{\rm LF}} $ and the projection operators $P_0$ and $Q_0$ commute with each other, we have
\begin{eqnarray} 
P_0 H_1P_0=H_{{\rm LF}}  \otimes  \ket{{\psi_0^{\md}}}\bra{{\psi_0^{\md}}}  + P_0 H_{{\rm couple}} P_0  
\end{eqnarray}
and
\begin{eqnarray} 
 P_0 H_1 Q_0 \frac{1}{E_0^\md - H_0}Q_0  H_1 P_0  =  P_0 H_{{\rm couple}} Q_0 \frac{1}{E_0^\md-H_0}Q_0  H_{{\rm couple}} P_0,
\end{eqnarray} 
where we utilized the equation $P_0 Q_0 = P_0(I_\tot-P_0)=0$. 
Then, we obtain the effective Hamiltonian for the spins~1 and $N$, which we refer to as $H_{1N}^\eff$, up to the constant $E_0^\md$, in the form
\begin{eqnarray} 
\fl H_{1N}^\eff \equiv \bra{{\psi_0^{\md}}} H^\eff   \ket{{\psi_0^{\md}}} = &H_{{\rm LF}}  +  \bra{{\psi_0^{\md}}} H_{{\rm couple}} \ket{{\psi_0^{\md}}}  \nonumber \\
 &+  \bra{{\psi_0^{\md}}} H_{{\rm couple}} Q_0 \frac{1}{E_0^\md-H_0}Q_0  H_{{\rm couple}}  \ket{{\psi_0^{\md}}} . \label{Effective_Hamiltonian_green_1N}
\end{eqnarray} 
Note that the effective Hamiltonian can be divided into the part which includes only $H_{\rm LF}$ and the part which includes only $H_{\rm couple}$.
The expression~(\ref{Effective_Hamiltonian_green_1N})  is essentially the same as the one derived in Ref.~\cite{LDE3}.

In order to relate the effective Hamiltonian~(\ref{Effective_Hamiltonian_green_1N}) to the argument in Section~\ref{Section_General formulation of the long-distance entanglement}, we prove the following equation 
\begin{eqnarray}
\tr_{1N} \frac{1}{Z_\tot} e^{-\beta H_\tot}=  \frac{1}{Z_\tot} e^{-\beta H_{1N}^\eff}   + O\bigl(\| H_1\|^2 \bigr)    \label{precision_effective1}
\end{eqnarray}
in the low-temperature limit $\beta\to\infty$, where $\tr_{1N}$ denotes the trace operation on the system except the probe spins~1 and $N$.
Equation~(\ref{precision_effective1}) means that the effective Hamiltonian defined in Eq.~(\ref{Effective_Hamiltonian_green_1N}) gives the density matrix up to the second order.
The proof is given as follows.
First, Eq.~(\ref{Green_function_effective122}) is Laplace-transformed to the following equation:
\begin{eqnarray}
P_0 e^{-\beta H_\tot}   P_0   =e^{-\beta H^\eff}   . \label{precision_effective2}
\end{eqnarray}
Because of the inequality (\ref{Condition_enough_small}),  perturbation theory yields
\begin{eqnarray}
\fl \frac{1}{Z_\tot} e^{-\beta H_\tot} &=\frac{1}{Z_\tot}  (P_0+Q_0) e^{-\beta H_\tot} (P_0+Q_0)  \nonumber \\
\fl&=  \frac{1}{Z_\tot}P_0 e^{-\beta H_\tot} P_0+ O\bigl(\| H_1\| \bigr) (P_0 A_1 Q_0 + Q_0A_1^\dagger P_0) +O\bigl(\| H_1\|^2 \bigr)  , \label{precision_effective3}
\end{eqnarray}
in the low-temperature limit $\beta\to\infty$, where we define $A_1$ as an $O(1)$ operator of the total system.
We can obtain Eq.~(\ref{precision_effective1}) from Eqs.~(\ref{precision_effective2}) and (\ref{precision_effective3}).

Let us characterize each term of the effective Hamiltonian $H_{1N}^\eff$ in Eq.~(\ref{Effective_Hamiltonian_green_1N}).
The first term in Eq.~(\ref{Effective_Hamiltonian_green_1N}) is the local fields on the probe spins, whereas the second term gives the effective fields 
of the order of $\|H_1\|$; the second term does not generate the interaction term because $H_{{\rm couple}}$ given in Eq.~(\ref{Definition_of_interaction_Hamiltonian}) does not include the terms such as $\sigma_1^i \otimes \sigma_N^j$ ($i,j=x,y,z$).
The third term gives the effective interaction between the probe spins as well as the effective fields of order $\|H_1\|^2$. 
The effective Hamiltonian $H_{1N}^\eff$ is thereby summarized as follows:
\begin{eqnarray}
H_{1N}^\eff =H_{{\rm LF}}+ \sum_{i=x,y,z} (h_1^{i,\eff}\sigma_1^i+ h_N^{i,\eff}\sigma_N^i) + \sum_{i,j=x,y,z}J_\eff^{i,j}\sigma_1^i\otimes\sigma_N^j,
\end{eqnarray}
where
\begin{eqnarray}
& \sum_{i=x,y,z}( h_1^{i,\eff}\sigma_1^i+ h_N^{i,\eff} \sigma_N^i) =   \bra{{\psi_0^{\md}}} H_{{\rm couple}} \ket{{\psi_0^{\md}}}  + O\bigl(\|H_1\|^2\bigr) , \nonumber \\
& \sum_{i,j=x,y,z}J_\eff^{i,j}\sigma_1^i\otimes\sigma_N^j = O\bigl(\|H_1\|^2\bigr).  \label{effective Hamiltonian365}
\end{eqnarray}

We show that a necessary condition for the mediator system to generate the long-distance entanglement is that the fields of  the order  of $\|H_1\|$ vanish, namely,
\begin{eqnarray} 
\bra{{\psi_0^{\md}}} H_{{\rm couple}} \ket{{\psi_0^{\md}}} +H_{{\rm LF}}=0.  \label{Condition_effective_field_zero}
\end{eqnarray}
If the above condition is not satisfied, the effective fields of order $O(\|H_1\|)$ totally destroy the long-distance entanglement in the limit of (\ref{the long-distance entanglement_generation}).
When we explicitly know the form of the effective Hamiltonian, we can cancel the effective fields by choosing the local fields on the spins~1 and $N$ as
\begin{eqnarray}
H_{{\rm LF}}= \sum_{i=x,y,z}\bigl( h_{1}^{i} \sigma_1^i   + h_{N}^{i}  \sigma_N^j \bigr)   = -  \bra{{\psi_0^{\md}}} H_{{\rm couple}} \ket{{\psi_0^{\md}}}.  
\end{eqnarray}
Note that the condition~(\ref{Condition_effective_field_zero}) does not ensure the maximum long-distance entanglement, namely $C(\rho_{1N})=1$; the effective fields of the order of $\|H_1\|^2$ can still exist even if the condition~(\ref{Condition_effective_field_zero}) is satisfied, which are of the same order as the indirect interaction in the limit of (\ref{the long-distance entanglement_generation}).

Let us consider the following special case:
\begin{eqnarray}
H_{{\rm couple}}=\sum_{i,j=x,y,z}\bigl(J_{A}^{i,j} \sigma_1^i  \otimes  \sigma_2^j   +J_{B}^{i,j} \sigma_N^i  \otimes  \sigma_{N-1}^j \bigr)
\end{eqnarray}
with the condition
\begin{eqnarray} 
\bra{{\psi_0^{\md}}} H_{{\rm couple}} \ket{{\psi_0^{\md}}} =0.  \label{Condition_effective_field_zero_SP}
\end{eqnarray}
In this case, we do not need $H_{\rm LF}$ to satisfy Eq.~(\ref{Condition_effective_field_zero}), namely we can put $H_{\rm LF}=0$.
The condition~(\ref{Condition_effective_field_zero_SP}) is reduced to
\begin{eqnarray} 
\fl \sum_{j=x,y,z}  \bra{{\psi_0^{\md} }} J_{A}^{i,j} \sigma_2^j  \ket{{\psi_0^{\md} }}=0\quad {\rm and} \quad \sum_{j=x,y,z}   \bra{{\psi_0^{\md} }} J_{B}^{i,j} \sigma_{N-1}^j  \ket{{\psi_0^{\md} }} =0      \label{Condition_effective_field_zero2exa}
\end{eqnarray}
for $i=x,y,z$.
This indeed occurs in any systems with the time-reversal symmetry; quantum spin chains without external fields and other odd-body interactions typically satisfy Eq.~(\ref{Condition_effective_field_zero2exa}), even if the Dzyaloshinskii-Moriya interaction~\cite{Dzyaloshinskii,Moriya} is included in the spin-spin interaction.

In the derivation~(\ref{Effective_Hamiltonian_green_1N}) of the effective Hamiltonian, we assumed that the effective Hamiltonian can be obtained by the second-order perturbation.
However, there are cases in which we cannot obtain  the effective interaction by the second-order perturbation.
For example, in the cases where the indirect interaction is `classical,' we cannot obtain the effective interaction by the second order; if the effective Hamiltonian for a classical interaction would be given as in Eq.~(\ref{Effective_Hamiltonian_green_1N}), we could always obtain the maximum entanglement by choosing the local fields $\{h_{1}^i,h_{N}^i\}_{i=x,y,z}$ properly~\cite{Kuwahara1}, but this is contradictory to Theorem~1.    
If the second-order perturbation vanishes, the effective interaction may depend on the local fields $\{h_{1}^i,h_{N}^i\}_{i=x,y,z}$ and takes a more complicated form in higher-order perturbations.



\section{The generation of the long-distance entanglement  by the local fields} \label{section_the generation of the long-distance entanglement  by the local fields}

In the present section, we discuss the generation of the long-distance entanglement  by the use of only the local fields.
Throughout the present section, we consider the system~(\ref{the long-distance entanglement_Hamiltonian_maximum_field}) in Fig.~\ref{fig:LDE_generation} (b).

\subsection{Effective Hamiltonian}
We can realize the condition equivalent to~(\ref{the long-distance entanglement_generation}) in the system (\ref{the long-distance entanglement_Hamiltonian_maximum_field}) by increasing the amplitudes of the local fields $\{h_2^{i}\}_{i=x,y,z}$ and $\{h_{N-1}^{i}\}_{i=x,y,z}$ with the local fields $\{h_1^{i}\}_{i=x,y,z}$ and $\{h_{N}^{i}\}_{i=x,y,z}$ canceling the resulting effective fields.
In the limits $|\vec{h}_2|\to \infty$ and $|\vec{h}_{N-1}|\to \infty$, the total Hamiltonian~(\ref{the long-distance entanglement_Hamiltonian_maximum_field}) can be transformed into the following effective Hamiltonian after tracing out the spins~2 and $N-1$:
\begin{eqnarray}
H_{\tot}^\eff=& \sum_{i,j=x,y,z}\bigl( J_{A}^{i,j,\eff} \sigma_1^i  \sigma_3^j  +J_{B}^{i,j,\eff} \sigma_{N-2}^i  \sigma_{N}^j \bigr)  \nonumber \\
&+\sum_{i=x,y,z} \bigl[ (h_1^{i} + h_1^{i,\eff}) \sigma_1^i +  (h_N^{i} + h_N^{i,\eff})  \sigma_N^i  \bigr]  +H_{{\rm media}}+H_{{\rm media}}^\eff .   \label{Kuwahara_Hamiltonian_effecitive}
\end{eqnarray} 
The coupling parameters  $\{J_{A}^{i,j,\eff}\}_{i,j=x,y,z}$ and $\{J_{B}^{i,j,\eff}\}_{i,j=x,y,z}$ in Eq.~(\ref{Kuwahara_Hamiltonian_effecitive}) approach zero as $|\vec{h}_2|^{-1}$ and $|\vec{h}_{N-1}|^{-1}$.
The amplitudes of the effective fields $\{h_1^{i,\eff}\}_{i=x,y,z}$ and $\{h_N^{i,\eff}\}_{i=x,y,z}$ are also of the order of $|\vec{h}_2|^{-1}$ and $|\vec{h}_{N-1}|^{-1}$, respectively, which is the same as the coupling parameters $\{J_{A}^{i,j,\eff}\}_{i,j=x,y,z}$ and $\{J_{B}^{i,j,\eff}\}_{i,j=x,y,z}$.
In order to achieve the condition~(\ref{Condition_effective_field_zero}), we  apply the weak local fields $\vec{h}_1=-\vec{h}_{1}^{\eff}$ and $\vec{h}_N=-\vec{h}_{N}^{\eff}$ on the spins~1 and $N$ to cancel the effective fields, so that
\begin{eqnarray}
&\sum_{i=x,y,z}  (h_1^{i} + h_1^{i,\eff}) \sigma_1^i =0 ,\nonumber \\
&\sum_{i=x,y,z} (h_N^{i} + h_N^{i,\eff})  \sigma_N^i   =0.   
\end{eqnarray}

In the following, for simplicity, we let the coupling parameters $\{J_{A}^{i,j},\tilde{J}_{A}^{i,j},\tilde{J}_{B}^{i,j},J_{B}^{i,j}\}_{i,j=x,y,z}$ be the \textit{XYZ} interaction and $\{h_2^{i},h_{N-1}^{i}\}_{i=x,y,z}$ applied only in the $z$ direction:
\begin{eqnarray}
H_{\tot}=& \sum_{i=x,y,z}\bigl( J_{A}^{i} \sigma_1^i  \sigma_2^i  + \tilde{J}_{A}^{i} \sigma_2^i \sigma_3^i   +\tilde{J}_{B}^{i} \sigma_{N-2}^i  \sigma_{N-1}^i  +J_{B}^{i} \sigma_{N-1}^i  \sigma_N^i \bigr)  \nonumber \\
&+\sum_{i=x,y,z} \bigl( h_1^{i} \sigma_1^i +  h_N^{i} \sigma_N^i  \bigr) +  h_2^{z} \sigma_2^z  +  h_{N-1}^{z} \sigma_{N-1}^z +H_{{\rm media}} .   \label{the long-distance entanglement_Hamiltonian_maximum_field_second}
\end{eqnarray} 
In  Section~\ref{The equivalence between the attenuation}, we show the decrease of the effective coupling interaction mathematically and discuss how precisely we can achieve the condition~(\ref{the long-distance entanglement_generation}) by increasing $|\vec{h}_2|$ and $|\vec{h}_{N-1}|$.


\subsection{The equivalence between the increase of the local fields  and  the decrease of the interaction} \label{The equivalence between the attenuation}
\begin{figure}
\centering
\includegraphics[clip, scale=0.6]{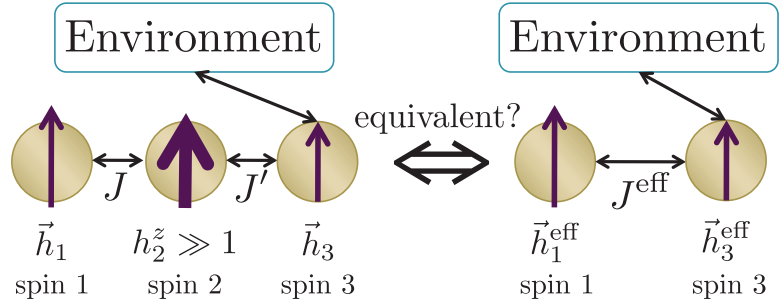}
\caption{Schematic picture on the equivalence between the decrease of the interaction and the increase of the local fields.
We can effectively weaken the indirect interaction between the spins~1 and 3 by increasing the local field on the spin~2. }
\label{fig:Equivalence_field_interaction}
\end{figure}

We here show the equivalence between the increase of the local fields  and  the decrease of the interaction.
In order to study the mathematical structure generally, we consider the quantum system shown in Fig.~\ref{fig:Equivalence_field_interaction} instead of the system in Fig.~\ref{fig:LDE_generation} (b); the spins~1 and 3 indirectly interact with each other through the spin~2, while the spin~3  is coupled to the environmental system.
We define the environment as an arbitrary $n_{{\rm env}}$-dimensional quantum system; that is, the dimensionality of the total system is $8n_{{\rm env}}$.
The total Hamiltonian is given as follows:
\begin{eqnarray}
&H_{\tot}=  h_2^z \sigma_2^z +H_{J,J'}  +H_{{\rm 3,env}},  \nonumber \\
&H_{J,J'}\equiv  \sum_{i=x,y,z}\bigl( J^{i} \sigma_1^i \sigma_2^i  + J^{'i} \sigma_2^i  \sigma_3^i\bigl),                  \label{Total_Hamiltonian_before_Effective} 
\end{eqnarray} 
where $H_{{\rm 3,env}}$ is an arbitrary $2n_{{\rm env}}$-dimensional Hamiltonian of the spin~3 and the environment.
The effective Hamiltonian $H_{{\rm 1,3,env}}^{\eff}$ after tracing out the spin~2  takes a simple form in the limit $|h_2^z|\to \infty$.   
We apply the formula~(\ref{Effective_Hamiltonian_green}) to the Hamiltonian~(\ref{Total_Hamiltonian_before_Effective}) to  obtain the form of $H_{{\rm 1,3,env}}^{\eff}$.

\textit{Theorem~2}:
The effective Hamiltonian $H_{{\rm 1, 3,env}}^{\eff}$ for the Hamiltonian~(\ref{Total_Hamiltonian_before_Effective}) is given as follows in the limit of $|h_2^z| \to \infty$:
\begin{eqnarray}
H_{{\rm 1,3,env}}^{\eff}=&\sum_{i=x,y,z} J^{i,\eff} \sigma_1^i  \sigma_3^i + H_{{\rm 3,env}} +h_{1}^{z,\eff} \sigma_1^z +h_{3}^{z,\eff} \sigma_3^z,   \label{effective_interaction_parameter_theorem1}
\end{eqnarray} 
where 
\begin{eqnarray}
&J^{x,\eff} =-\frac{J^xJ^{'x}}{h_2^z}  ,\quad J^{y,\eff} =-\frac{J^yJ^{'y}}{h_2^z} ,\quad  J^{z,\eff}=0 ,\nonumber \\
&h_{1}^{z,\eff}=-J^z -\frac{J^xJ^y}{h_2^z}, \quad  h_{3}^{z,\eff}=-J^{'z} -\frac{J^{'x}J^{'y}}{h_2^z}  . \label{correspondence_effective_real}
\end{eqnarray}
In addition, the Hamiltonian $H_{{\rm 1,3,env}}^{\eff}$ satisfies
\begin{eqnarray}
\tr^2 \frac{ e^{-\beta H_{\tot}}}{Z_\tot}= \frac{e^{-\beta H_{{\rm 1,3,env}}^{\eff}}}{Z_\tot} + O(|h_2^z|^{-2})  \label{the_degree_of_precision_theorem1}
\end{eqnarray}
in the limit of $\beta\to\infty$,
where $\tr^{2}$ denotes the trace operation only on the spin~2.

\textit{Comment}:
The influence from the spin~2 is effectively expressed as the fields on the spins~1 and 3 after the trace out of the spin~2.
The interaction with the spin~2 also mediates the indirect interaction between the spins~1 and 3.
As the external fields $|h_2^z|$ increases, the state of the spin~2 is approximately fixed to $\ket{\ua_2}$, and hence, the indirect interaction via the spin~2 is weakened.
The indirect interactions along the $x$ and $y$-axes decay as $|h_2^z|^{-1}$ in the limit $|h_2^z|\to\infty$, while the indirect interaction in the $z$-axis decays as $|h_2^z|^{-t}$ with $t \ge2$. 
In other words, the indirect interaction in the direction of the local field on the spin~2 decays more rapidly.


\textit{Proof}:
In order to prove the present theorem, we  follow the same calculations as in Section~\ref{Section_Effective_field}.   
We break down the Hamiltonian~(\ref{Total_Hamiltonian_before_Effective}) as follows:
\begin{eqnarray}
&H_{\tot}=H_0 + H_1 ,\quad H_0\equiv  h_2^z \sigma_2^z, \nonumber \\
&H_1\equiv H_{J,J'} + H_{{\rm 3,env}} .    \label{Hamiltonian_separate_for_perturbation}
\end{eqnarray} 
We define the ground state of $H_0$ as $\ket{{\ua_2}}$ with the eigenvalue $-h_2^z$.

First, we prove Eq.~(\ref{effective_interaction_parameter_theorem1}).
We can calculate the effective Hamiltonian as in Eq.~(\ref{Effective_Hamiltonian_green}):
\begin{eqnarray} 
H^\eff =P_0 H_0 P_0+P_0 H_1P_0 + P_0 H_1 Q_0 \frac{1}{-h_2^z-H_0}Q_0  H_1 P_0  \label{Effective_Hamiltonian_green_LDE},
\end{eqnarray}
where $P_0\equiv \ket{{\ua_2}} \bra{{\ua_2}}\otimes I_{1,3,\env}$ with $I_{1,3,\env}$ the identity matrix of the spins~1, 3 and the environment, and $Q_0\equiv I_\tot - P_0$.
Because $H_{{\rm 3,env}}$ commutes with $P_0$, we have
\begin{eqnarray} 
\fl H^\eff =\ket{{\ua_2}} \bra{{\ua_2}}\otimes  \biggl[  -h_2^z + H_{{\rm 3, env}} +\bra{{\ua_2}} \Bigl( H_{J,J'}+H_{J,J'} Q_0 \frac{1}{-h_2^z -H_0}Q_0  H_{J,J'}\Bigr) \ket{{\ua_2}} \biggr].  \label{Effective_Hamiltonian_green_LDE2}
\end{eqnarray}
The effective Hamiltonian $H_{{\rm 1,3,env}}^{\eff}$ of the spins 1, 3 and the environment is thereby given by $H_{{\rm 1,3,env}}^{\eff}=\bra{{\ua_2}}H^\eff \ket{{\ua_2}}$.
We can calculate the term $\bra{{\ua_2}} \Bigl( H_{J,J'}+H_{J,J'} Q_0 \frac{1}{-h_2^z -H_0}Q_0  H_{J,J'}\Bigr) \ket{{\ua_2}}$ as follows:
\begin{eqnarray} 
\fl &\bra{{\ua_2}}  H_{J,J'}  \ket{{\ua_2}} +\bra{{\ua_2}} H_{J,J'} Q_0 \frac{1}{-h_2^z -H_0}Q_0  H_{J,J'}  \ket{{\ua_2}}  \nonumber \\
\fl & = (-J^z \sigma_1^z -J^{'z} \sigma_3^z)+ \biggl(  -\frac{J^xJ^y}{h_2^z} \sigma_1^z  -\frac{J^{'x}J^{'y}}{h_2^z} \sigma_3^z -\frac{J^xJ^{'x}}{h_2^z} \sigma_1^x  \sigma_3^x -\frac{J^yJ^{'y}}{h_2^z}\sigma_1^y  \sigma_3^y \biggr).
\end{eqnarray}
From the above calculation, we arrive at Eq.~(\ref{effective_interaction_parameter_theorem1}) with Eq.~(\ref{correspondence_effective_real}) up to the constant component $-h_2^z$.

Next, we prove Eq.~(\ref{the_degree_of_precision_theorem1}). 
Assuming $h_2^z \gg \|H_1\|$, we obtain 
\begin{eqnarray}
\frac{ e^{-\beta H_{\tot}}}{Z_\tot} =P_0 \frac{ e^{-\beta H_{\tot}}}{Z_\tot} P_0 +O( |h_2^z|^{-1})( P_0 A_1Q_0+Q_0 A_1^\dagger P_0)+ O( |h_2^z|^{-2}) \label{precision_effective_fields_section33}
\end{eqnarray}
in the low-temperature limit as in Eq.~(\ref{precision_effective3}), where we define  $A_1$ as an $O(1)$ operator of the total system.
We then obtain Eq.~(\ref{the_degree_of_precision_theorem1}) from Eqs.~(\ref{precision_effective2}) and (\ref{precision_effective_fields_section33}).
This completes the proof of Theorem~2. $\opensquare$


Applying Theorem~2 to the Hamiltonian~(\ref{the long-distance entanglement_Hamiltonian_maximum_field}), we obtain the effective Hamiltonian $H_{\tot}^{\eff}$ of the form~(\ref{Kuwahara_Hamiltonian_effecitive}).
For the Hamiltonian~(\ref{the long-distance entanglement_Hamiltonian_maximum_field_second}), more specifically, the effective Hamiltonian $H_{\tot}^{\eff}$ is given by
\begin{eqnarray}
H_{\tot}^{\eff}=&H_{\rm LF}+H_{\rm LF}^{\eff} +  H_{{\rm couple}}^{\eff}     +H_{{\rm media}}+H_{{\rm media}}^\eff   ,     \label{effective_Hamiltonian_for_calculation}
\end{eqnarray} 
where
\begin{eqnarray}
&H_{\rm LF}^{\eff} =-\biggl(J_A^z +\frac{J_A^xJ_A^y}{h_2^z}\biggr)\sigma_1^z -\biggl( J_B^z +\frac{J_B^x J_B^y}{h_{N-1}^z}\biggr) \sigma_N^z \nonumber \\
&H_{{\rm couple}}^{\eff}\equiv -\frac{J_A^xJ_A^{'x}}{h_2^z} \sigma_1^x \sigma_3^x  -\frac{J_A^y J_A^{'y}}{h_2^z} \sigma_1^y\sigma_3^y   - \frac{J_B^xJ_B^{'x}}{h_{N-1}^z} \sigma_{N-2}^x \sigma_N^x  -\frac{J_B^y J_B^{'y}}{h_{N-1}^z}\sigma_{N-2}^y \sigma_N^y  ,    \nonumber \\
&H_{{\rm media}}^\eff \equiv -\biggl(\tilde{J}_A^z +\frac{\tilde{J}_A^x\tilde{J}_A^y}{h_2^z}\biggr) \sigma_3^z -\biggl (\tilde{J}_B^z +\frac{\tilde{J}_B^x \tilde{J}_B^y}{h_{N-1}^z}\biggr)\sigma_{N-2}^z  .
\end{eqnarray} 
We thus achieve $\|H_{{\rm couple}}\|\to0$ in the limits $|h_2^z|\to \infty$ and $|h_{N-1}^z|\to \infty$.
We can choose the local fields $h_1^z$ and $h_{N}^z$ as
\begin{eqnarray}
h_1^z=J_A^z +\frac{J_A^xJ_A^y}{h_2^z},\quad h_{N}^z=J_B^z +\frac{J_B^x J_B^y}{h_{N-1}^z}       \label{effective_fields_for_calculation}
\end{eqnarray}
so that they may cancel the effective fields $H_{\rm LF}^{\eff}$.

\subsection{Numerical demonstration}  \label{Sec:Numerical_calculation_LDE}
In the present subsection, we numerically demonstrate the generation of the long-distance entanglement by the use of the local fields for an $XY$ spin chain.
We consider the total Hamiltonian
\begin{eqnarray}
H_{\tot}=&(1+\gamma)( \sigma_1^x   \sigma_2^x + \sigma_2^x   \sigma_3^x +     \sigma_{N-2}^x   \sigma_{N-1}^x +  \sigma_{N-1}^x   \sigma_N^x) \nonumber \\
 &+ (1- \gamma)( \sigma_1^y   \sigma_2^y + \sigma_2^y  \sigma_3^y +     \sigma_{N-2}^y   \sigma_{N-1}^y +  \sigma_{N-1}^y   \sigma_N^y)  \nonumber \\
&+\frac{(1-\gamma^2)}{h_0} (\sigma_1^z+\sigma_N^z )+h_0 ( \sigma_2^z+ \sigma_{N-1}^z)+H_{\md}, \nonumber \\
H_{\md}=&\sum_{l=3}^{N-3} \bigl[(1+\gamma) \sigma_l^x   \sigma_{l+1}^x + (1- \gamma) \sigma_l^y   \sigma_{l+1}^y \bigr] +\frac{(1-\gamma^2)}{h_0} (\sigma_3^z+\sigma_{N-2}^z ) , \label{XYmodel_the long-distance entanglement1}
\end{eqnarray}
where we choose the local fields on the spins~2 and $N-1$ as $h_2^z=h_{N-1}^z=h_0$.
We introduced the  third term in $H_{\tot}$ accordingly to Eq.~(\ref{correspondence_effective_real})  in order to cancel the effective fields on the spins~1 and $N$ which are generated after tracing out the spins~2 and $N-1$.
We also introduced the second term in $H_\md$ to cancel the effective fields on the media spins~3 and $N-2$ for simplicity.
We obtain the effective Hamiltonian $H_{\tot}^{\eff}$ of the form~(\ref{effective_Hamiltonian_for_calculation}) as
\begin{eqnarray}
H_{\tot}^{\eff}=&-\frac{(1+\gamma)^2}{h_0} (\sigma_1^x\sigma_3^x +\sigma_{N-2}^x\sigma_N^x)-  \frac{(1-\gamma)^2}{h_0} (\sigma_1^y\sigma_3^y+\sigma_{N-2}^y\sigma_N^y) \nonumber  \\
&+\sum_{l=3}^{N-3} \bigl[(1+\gamma) \sigma_l^x   \sigma_{l+1}^x+ (1- \gamma) \sigma_l^y   \sigma_{l+1}^y \bigr]  \label{XYmodel_effective_LDE}
\end{eqnarray}
in the limit of $h_0\to\infty$.
The ground state of the $XY$ spin chain without fields satisfies the condition~(\ref{Condition_effective_field_zero_SP}) because the Hamiltonian has the time-reversal symmetry.
Note that the indirect interaction through the $XY$ spin chain is not a `classical' one.
We can therefore expect that the long-distance entanglement exists in this system in the limit of $h_0\to\infty$. 
  
  \begin{figure}
\centering
\subfigure[]{
\includegraphics[clip, scale=0.7]{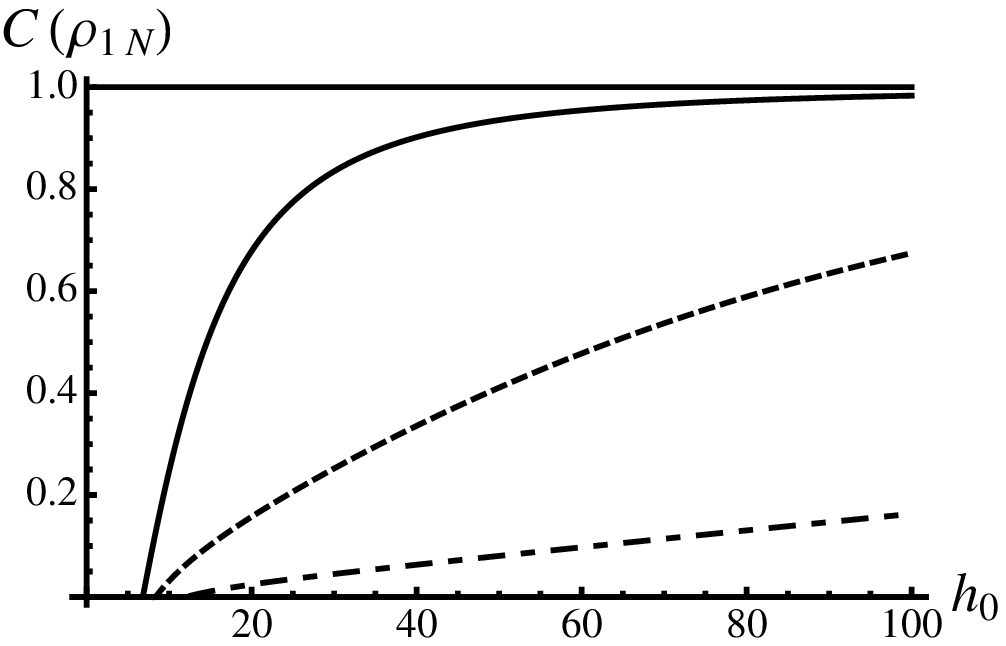}
}
\subfigure[]{
\includegraphics[clip, scale=0.7]{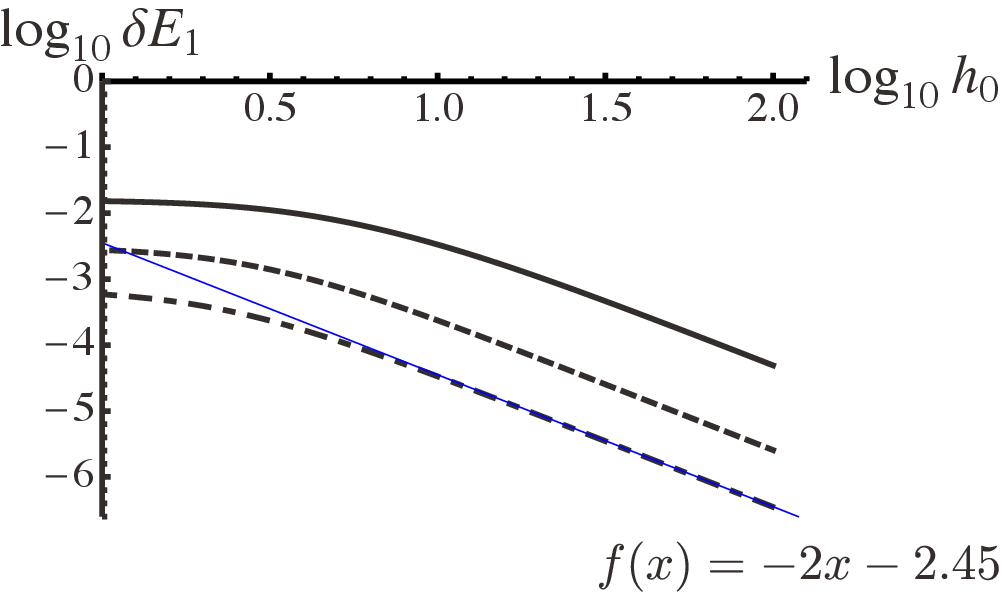}
}
\caption{
We plot (a) the long-distance entanglement $C(\rho_{1N})$ and (b) the first excitation energy $\delta E_1$ against the field $h_0$ in the three cases $\gamma=0$ (solid line), $\gamma=0.03$ (broken line) and $\gamma=0.05$ (chained line). The fitting line (thin solid line) to the data points shows that the first excitation energy decreases as $h_0^{-2}$.
}
\label{fig:the long-distance entanglement_XX_XY}
\end{figure}

\begin{table}[t]
\begin{center}
\begin{tabular}{cccc}
\hline
 $\gamma$ & 0 & 0.03& 0.05 \\  \hline
\ \ Hamiltonian~(\ref{XYmodel_the long-distance entanglement1})\ \  & $\ 0.9832\ $ & $\ 0.6743\ $ & $\ 0.1631\ $  \\
\ \ Hamiltonian~(\ref{XYmodel_effective_LDE})\ \ &$\ 0.9834\ $ & $\ 0.6745\ $ & $\ 0.1632\ $   \\ 
\hline
\end{tabular}
\caption{Comparison of the entanglement for $h_0=100$. The effective Hamiltonian of the total Hamiltonian (\ref{XYmodel_the long-distance entanglement1}) is equal to the Hamiltonian~(\ref{XYmodel_effective_LDE}) by the second-order perturbation. The difference between them is of the order of $h_0^{-2}=10^{-4}$.}
\label{Table:Comparison}
\end{center}
\end{table}

Let us show the numerical calculation of the long-distance entanglement for the system (\ref{XYmodel_the long-distance entanglement1}) with $N=100$.    
In Fig.~\ref{fig:the long-distance entanglement_XX_XY}~(a), we show the entanglement between the spins~1 and $N$ in the three cases, $\gamma=0$, $0.03$ and $0.05$.
We can see that the entanglement $C(\rho_{1N})$ monotonically  increases with $h_0$.
As the parameter $\gamma$ increases, however, we need a greater value of $h_0$ to generate a large long-distance entanglement.
This is because the first excitation energy of the mediator system decreases rapidly as the parameter $\gamma$ increases; as is shown in Eq.~(\ref{Condition_enough_small}), we need to attenuate the coupling strength adequately so that it may be much less than the first energy gap of the mediator system. 
If the condition (\ref{Condition_enough_small}) does not hold, the second term on the right-hand side of Eq.~(\ref{precision_effective1}) may not be ignored and may decrease the purity of the total system, which causes the destruction of the long-distance entanglement.
We show in Fig.~\ref{fig:the long-distance entanglement_XX_XY}~(b) the first excitation energy $\delta E_1$ of the total system, which decreases as $h_0^{-2}$ in the limit of $h_0\to\infty$; this dependence comes from the fact that the indirect interaction obtained from Eq.~(\ref{Effective_Hamiltonian_green_1N}) is of the order of $h_0^{-2}$.
This decrease of the excitation energy would make the long-distance entanglement fragile to thermal fluctuation. 

The value of the long-distance entanglement is almost the same  if we consider the total Hamiltonian~(\ref{XYmodel_effective_LDE}) with $N=98$ from the beginning.
We compare the values in Table~\ref{Table:Comparison} for $h_0=100$. 
The difference of $C(\rho_{1N})$ is of the order of $h_0^{-2}$ as is expected from Eq.~(\ref{the_degree_of_precision_theorem1}).



%

%

%

%

%

%

%

\section{Finite coupling}  \label{the long-distance entanglement_numerical_calculation_example}
In the previous sections, we considered the weak coupling limit (\ref{the long-distance entanglement_generation}) in order to generate the long-distance entanglement.
By making the coupling strength to zero, however, the first excitation energy of the total system vanishes and the long-distance entanglement becomes extremely fragile against thermal fluctuation.
In order to avoid the situation, we have to achieve the long-distance entanglement by as strong coupling as possible. 
In the present section, we consider the Hamiltonian~(\ref{the long-distance entanglement_Hamiltonian_total}) in Fig.~\ref{fig:LDE_generation} (a) with a finite coupling Hamiltonian:
\begin{eqnarray}
\|H_{\cop}\| ={\rm const} \neq 0.
\end{eqnarray}
We focus on the following two points:
\begin{enumerate}
\item{}We look for the mediator system suitable for the generation of the long-distance entanglement.
\item{}We do not necessarily need the condition~(\ref{Condition_effective_field_zero}) for the generation of the long-distance entanglement when the coupling Hamiltonian is finite.
\end{enumerate}

First, in Section~\ref{The_enhancement_of_LDE }, we introduce two examples  of the $XY$ spin chains with the external fields in the $z$ direction in order to demonstrate that the external fields on the mediator system can enhance the long-distance entanglement with the coupling strength fixed.
With a finite coupling Hamiltonian,  the preparation of a suitable mediator system is crucial for the generation of the long-distance entanglement.
For example, we showed in Section~\ref{Sec:Numerical_calculation_LDE} that the long-distance entanglement strongly depends on the parameter $\gamma$ if we keep the finite coupling strength.
We here instead consider the possibilities that we can enhance the capability of the mediator system with external fields on the mediator system.

Second, in Section~\ref{Small spin chains with anisotropy}, we discuss the case in which a random Hamiltonian, which may generate the effective fields~(\ref{effective Hamiltonian365}), is added to the system.
There, we tune the coupling strength in order to achieve the maximum entanglement and see that the coupling strength should be neither too strong nor too weak.
In the limit of the weak coupling~(\ref{the long-distance entanglement_generation}) of this system, the long-distance entanglement always vanishes in any quantum systems which do not satisfy the condition~(\ref{Condition_effective_field_zero}).
If the coupling has a finite value, however, the entanglement may exist in the system without the condition~(\ref{Condition_effective_field_zero}).
As has been analyzed in Section~\ref{Section_Effective_field}, the effective fields on the probe spins are of the first order of the coupling amplitude $O(\|H_\cop\|)$.
On the other hand, the effective interaction is of the order of $\|H_\cop\|^2$, and hence the ratio of the effective fields to the effective interaction is of the order of $\|H_\cop\|^{-1}$.
We therefore expect that by increasing the coupling strength we can relatively reduce the effective fields and enhance the long-distance entanglement.

\subsection{Enhancement of the long-distance entanglement with the external fields}\label{The_enhancement_of_LDE }

In the present section, we consider the possibilities that we can enhance the capability of the mediator system to generate the long-distance entanglement by the use of the external fields on the mediator system.
We consider the $XY$ spin chains with the external fields in the $z$ direction in the following two cases: random fields and uniform fields.
First, it has been shown that in the $XX$ spin chain the random fields can enhance the entanglement between short-range spin pairs~\cite{Fujinaga}.
We thereby expect that a randomness may also enhance the long-distance entanglement.
Second, it is known that the quantum phase transition occurs in the $XY$ spin chains with the external fields.
Then, we expect that the long-distance entanglement is enhanced around the critical point because of strong quantum fluctuation.
We indeed show that the long-distance entanglement is highly enhanced in these two cases.

We first give the Hamiltonian
\begin{eqnarray}
H_{\tot}=&  H_\md+0.02 (\sigma_1^x   \sigma_{2}^x +\sigma_1^y   \sigma_{2}^y+\sigma_{N-1}^x   \sigma_{N}^x +\sigma_{N-1}^y   \sigma_{N}^y )  , \label{XYmodel_the long-distance entanglement2}
\end{eqnarray}
where
\begin{eqnarray}
H_\md=&  \sum_{l=2}^{N-2} \bigl[\alpha(1+\gamma) \sigma_l^x   \sigma_{l+1}^x + (1- \gamma) \sigma_l^y   \sigma_{l+1}^y \bigr] +\sum_{l=2}^{N-1} h_l^z \sigma_l^z
\end{eqnarray}
and $\alpha$ is an integer which has the values $\pm1$. 
The property of the mediator system qualitatively changes depending on the sign $\alpha$ as is shown in the following.
In the present system, the indirect interaction between the spins~1 and $N$ depends on the external fields $\{h_l^z\}_{l=2}^{N-1}$.
We then expect the possibilities that $\{h_l^z\}_{l=2}^{N-1}$ enhance the generation of the long-distance entanglement.
\begin{figure}
\centering
\subfigure[]{
\includegraphics[clip, scale=0.7]{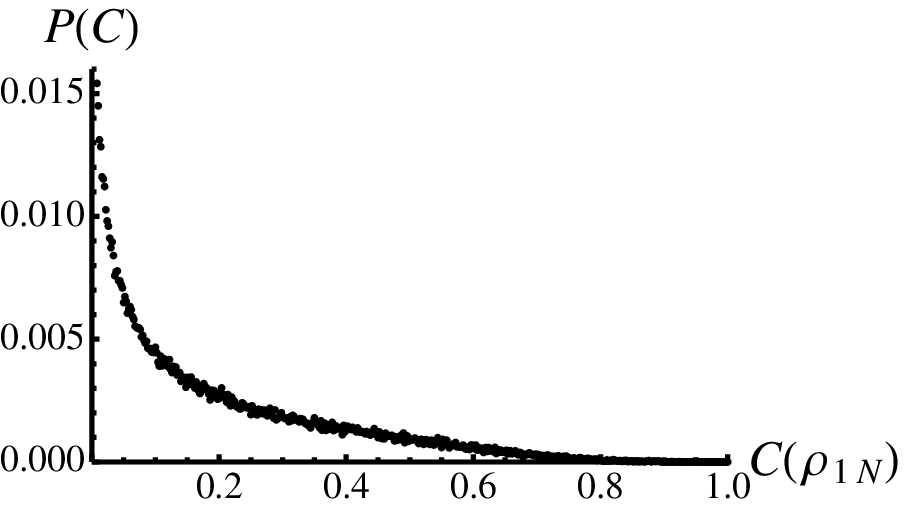}
}
\subfigure[]{
\includegraphics[clip, scale=0.7]{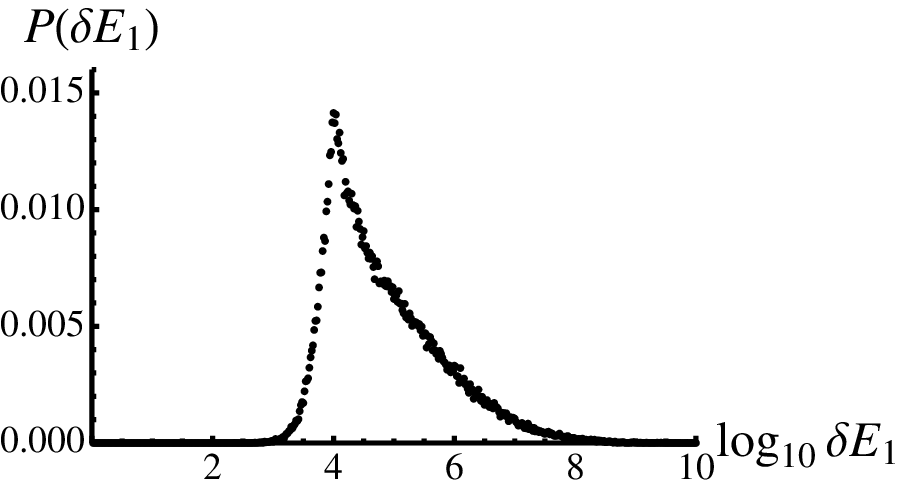}
}
\caption{The distribution of (a) the long-distance entanglement and (b) the first excitation energy. 
We determined each of the external fields $\{h_l\}_{l=2}^{99}$ stochastically out of the uniform distribution $[-1.5,1.5]$, to obtain $10^7$ samples. 
In (a), the value for the bin from 0 to 0.002 is  $0.134$, which is out of the range of the plot.
The average of the entanglement is $0.162$.
Without random fields, the value of the long-distance entanglement would be $C=1.69\times 10^{-4}$ and the first excitation energy would be $\delta E_1=1.19\times10^{-8}$.}
\label{fig:the long-distance entanglement_rand_h1.5}
\end{figure}

First, we apply the random fields $\{h_l^z\}_{l=2}^{N-1}$ distributed uniformly in the range $-1.5\le h^z_l\le1.5$ ($l=1,2,\ldots, N-1$). 
We let $\alpha=+1$, $\gamma=0.1$ and $N=100$.
For $h_l^z=0$ ($l=2,\ldots, 99$), the long-distance entanglement $C(\rho_{1N})$ and the first excitation energy $\delta E_1$ are equal to $1.69\times 10^{-4}$ and $1.19\times10^{-8}$, respectively.
In Fig.~\ref{fig:the long-distance entanglement_rand_h1.5}, we show the distributions of $C(\rho_{1N})$ and $\log_{10}\delta E_1$ in the case of the random fields.
The average of $C(\rho_{1N})$ over $10^7$ samples is $0.162$; we can see that the random fields  highly improve the entanglement and the first excitation energy.
There are even samples for which the entanglement $C(\rho_{1N})$ is more than $0.95$; the largest value of the entanglement among the $10^7$ samples is $0.982$.
This suggests that we can significantly enhance the long-distance entanglement by choosing the external fields optimally.

 \begin{figure}
\centering
\subfigure[$N=100$]{
\includegraphics[clip, scale=0.7]{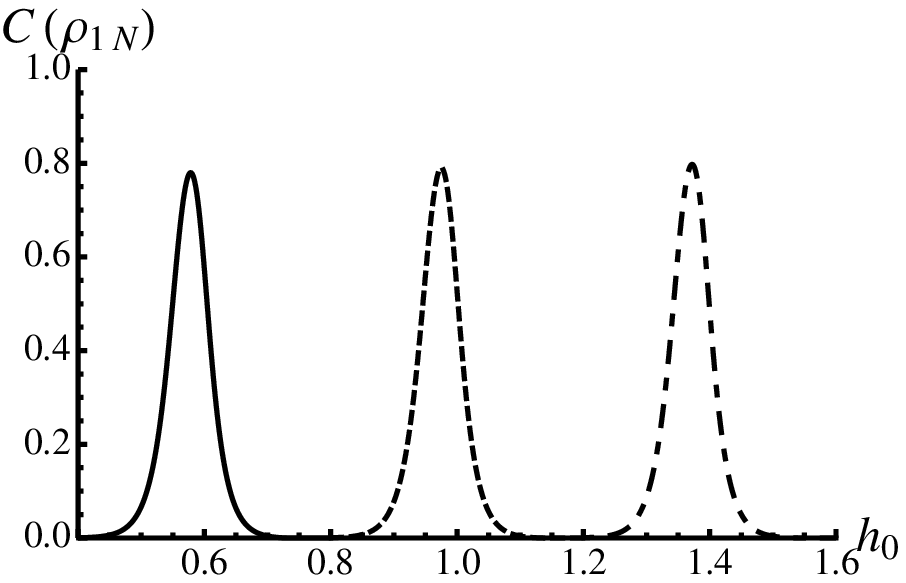}
}
\subfigure[$\gamma=0.5$]{
\includegraphics[clip, scale=0.7]{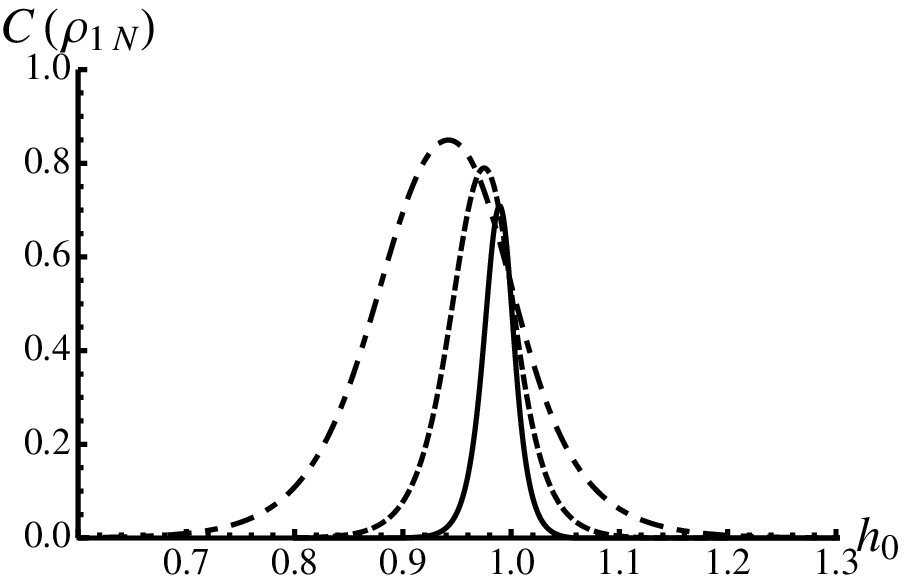}
}
\caption{The entanglement against $h_0$ (a) for $\gamma=0.3$ (solid line), $\gamma=0.5$ (broken line) and  $\gamma=0.7$ (chained line) with $N=100$ and (b) for  $N=50$ (chained line), $N=100$ (broken line) and $N=200$ (solid line) with $\gamma=0.5$. We apply the uniform fields $h_0$ on the \textit{XY} spin chain~(\ref{XYmodel_the long-distance entanglement2}). The long-distance entanglement can be enhanced around the point $h_0 = 2\gamma$. As the spin number $N$ increases, the peak approaches the critical point $h_0 = 2\gamma$ and becomes sharp.}
\label{fig:the long-distance entanglement_oscillational}
\end{figure}

Second, we apply the uniform fields in the case $\alpha=-1$:
\begin{eqnarray}
h_l^z=h_0
\end{eqnarray}
for $l=2,\ldots, N-1$.
In this case of $\alpha=-1$, a quantum phase transition occurs at the point $h_0 = 2\gamma$ in the limit of $N\to\infty$~\cite{XYquantum_phase_transition}.
In Fig.~\ref{fig:the long-distance entanglement_oscillational}~(a), we show the plots of the long-distance entanglement against the parameter $h_0$ for $\gamma=0.3,0.5,0.7$ and $N=100$.
We can see that the entanglement is highly enhanced near the critical point $h_0 = 2\gamma$.
In Fig.~\ref{fig:the long-distance entanglement_oscillational}~(b), we show for $\gamma=0.5$ that the peak becomes sharp and close to the point $h_0 = 2\gamma$ as the spin number $N$ increases from $50$ to $200$. 
For $\alpha=1$, we cannot achieve such enhancement by the external fields even at the critical point.

\subsection{Small spin chains with the random Hamiltonian} \label{Small spin chains with anisotropy}
 \begin{figure}
\centering
\subfigure[]{
\includegraphics[clip, scale=0.6]{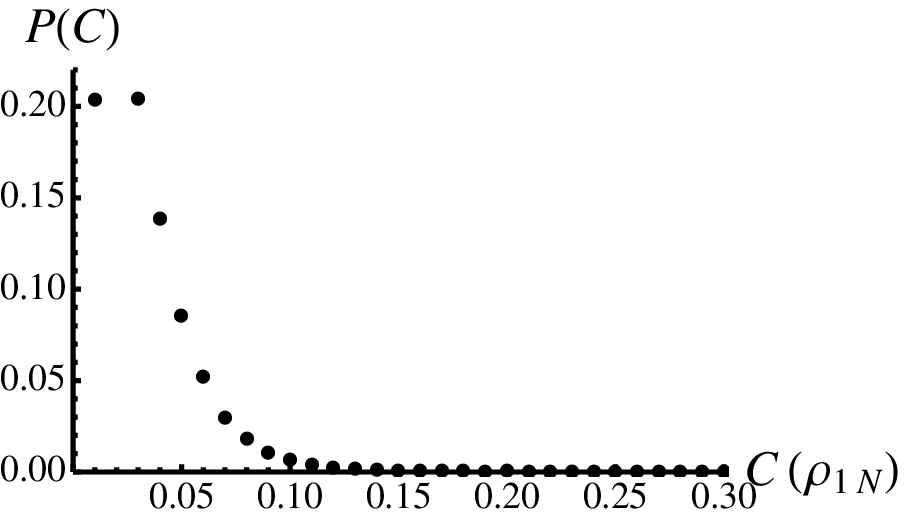}
}
\subfigure[]{
\includegraphics[clip, scale=0.6]{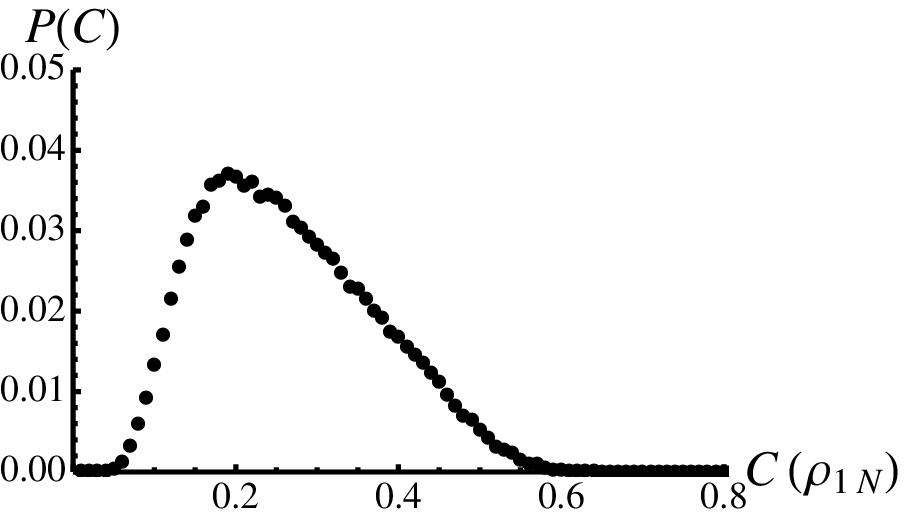}
}
\subfigure[]{
\includegraphics[clip, scale=0.6]{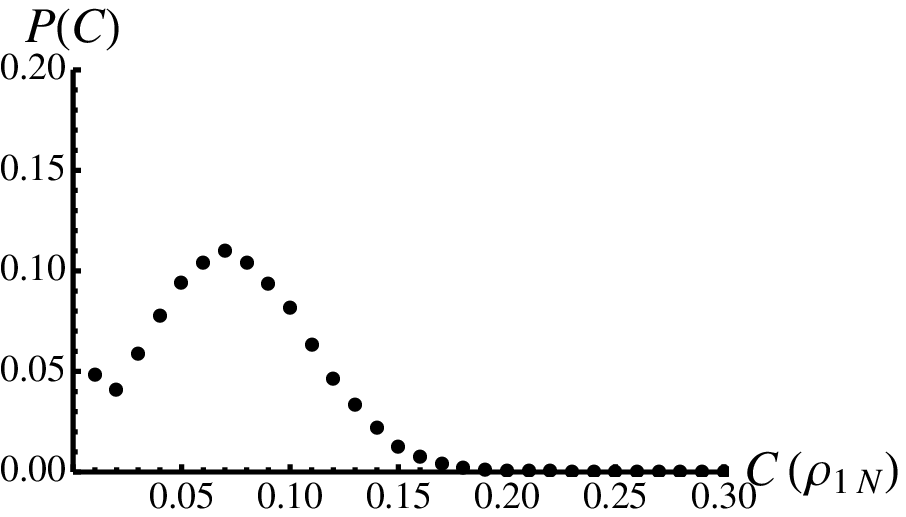}
}
\subfigure[]{
\includegraphics[clip, scale=0.6]{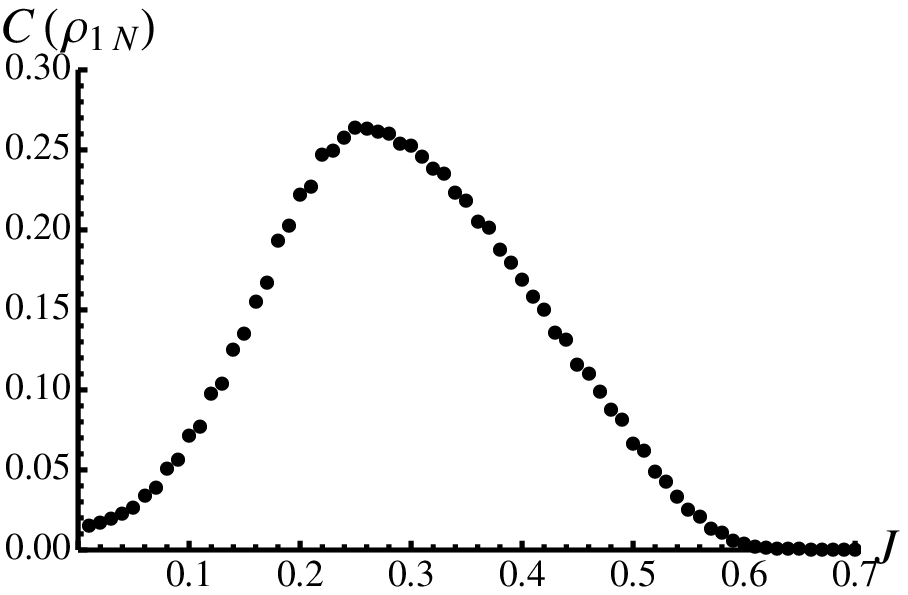}
}
\caption{The distribution of the entanglement $C(\rho_{1N})$ (a) for $J=0.05$, (b) for $J=0.25$ and (c) for $J=0.5$ for the Hamiltonian~(\ref{XXmodel_the long-distance entanglemen}).  
We determined each of the random coupling $\{\delta J_l^{i,j}\}_{i,j=0,x,y,z}$ ($l=1,2,\ldots,7$) stochastically out of the uniform distribution $[-0.05,0.05]$.
For the panels~(a)--(c), we used 250 000 samples for each case.
In  the panel~(d), we plot the average of $C(\rho_{1N})$ for each value of $J$ with 2500 samples for each point to calculate the average of the entanglement. 
}
\label{fig:the long-distance entanglement_Protect}
\end{figure}

Next, we consider a quantum system which does not satisfy the condition~(\ref{Condition_effective_field_zero}).
In order to discuss the effect of the breaking of the condition~(\ref{Condition_effective_field_zero}), we consider the following spin chain with $N=8$,
\begin{eqnarray}
H=&\sum_{l=2}^{6} \bigl(\sigma_l^x   \sigma_{l+1}^x + \sigma_l^y   \sigma_{l+1}^y \bigr)+J(\sigma_1^x   \sigma_{2}^x +\sigma_1^y   \sigma_{2}^y+\sigma_{7}^x   \sigma_{8}^x +\sigma_{7}^y   \sigma_{8}^y )    \nonumber \\
&+ \sum_{l=1}^{7} \sum_{i,j=0,x,y,z} \delta J_l^{i,j}\sigma_l^i   \sigma_{l+1}^j  , \label{XXmodel_the long-distance entanglemen}
\end{eqnarray}
where $\sigma^{0}$ denotes the identity operator.
We add the random coupling $\{\delta J_l^{i,j}\}_{i,j=0,x,y,z}$ ($l=1,2,\ldots,7$) to the \textit{XX} spin chains; we assume that the randomness is uniform in the range
$-0.05\le \delta J_l^{i,j} \le0.05$
for $i,j=0,x,y,z$ and $l=1,2,\ldots,7$.
In this system, the condition~(\ref{Condition_effective_field_zero}) is not satisfied, and hence the entanglement $C(\rho_{1N})$ is equal to zero in the limit $J\to0$.

\begin{table}[t]
\begin{center}
\begin{tabular}{cccc}
\hline
 $J$ & 0.05 & 0.25& 0.5 \\  \hline
 Entanglement without random Hamiltonian &$\ 0.973\ $& $\ 0.551\ $& $\ 0.102\ $  \\ 
Entanglement with random Hamiltonian & $0.03$& $0.27$& $0.07$ \\
\hline
\end{tabular}
\caption{The average values of the entanglement without and with the random Hamiltonian in Eq.~(\ref{XXmodel_the long-distance entanglemen}) }
\label{Table:Comparison2}
\end{center}
\end{table}

In Fig.~\ref{fig:the long-distance entanglement_Protect}~(a)--(c), we show the distribution of the entanglement $C(\rho_{1N})$ for $J=0.05$, $0.25$ and $0.5$.
In Table~\ref{Table:Comparison2}, we show the averages of the entanglement in these three cases.
We also show the values of the entanglement without the random Hamiltonian, namely $J_l^{i,j}=0$ for $i,j=0,x,y,z$ and $l=1,2,\ldots,7$.
We can see from Table~\ref{Table:Comparison2} that the destruction of the entanglement by the random Hamiltonian becomes smaller as the coupling parameter $J$ increases.
However, a too strong coupling parameter does not give the optimum value of the entanglement nor does a too weak coupling parameter. 
If the coupling parameter $J$ is small, the destruction of the entanglement by the effective fields is serious. 
On the other hand, if the coupling parameter $J$ is large, the entanglement between the probe spins and the mediator system becomes non-negligible, the purity of the probe spins  decreases, and hence the entanglement is destroyed.
The mechanism of the entanglement destruction is thus different physically in these two cases.
These two effects compete in giving the optimum value of the coupling parameter.
As is shown in Fig.~\ref{fig:the long-distance entanglement_Protect}~(d), we can achieve the maximum average of the entanglement with the coupling parameter $J\simeq0.25$.



%

%

\section{SUMMARY AND CONCLUSION}\label{conclusion}
We have analytically and numerically studied the generation of the long-distance entanglement by the use of the weak coupling~(\ref{the long-distance entanglement_generation}).
We gave the two necessary conditions for the mediator system to generate the long-distance entanglement.
The first one is that the indirect interaction must not be `classical' as has been defined in Eq.~(\ref{Sep_cond}).
The second one is that the effective fields of the first order have to vanish as in Eq.~(\ref{Condition_effective_field_zero}).
The first condition, in particular, is applicable to any bipartite systems indirectly interacting with each other at any temperatures.
The second condition can be satisfied artificially by applying the local fields so as to cancel the effective fields.
In this sense, we may overcome the constraint of the second condition if we explicitly know the effective fields on the probe spins.
We have shown that these two conditions can be satisfied in various systems; this means that many quantum systems have  potentials of the generation of the long-distance entanglement.

Next, we discussed the generation of the long-distance entanglement by the use of only the external fields.
As shown in Fig.~\ref{fig:LDE_generation} (b), we have to control the local fields on the probe spins and the part of the system adjacent to the probe spins.
We have to apply strong fields on the adjacent spins and the weak fields on the probe spins. 
Then, we achieved the condition mathematically equivalent to Eq.~(\ref{the long-distance entanglement_generation}).
The strong fields contribute to attenuation of the coupling between the probe spins and the mediator system, while weak fields are necessary to cancel the effective fields on the probe spins.
Because we utilize the fields with finite amplitudes, we can achieve the condition~(\ref{the long-distance entanglement_generation}) by the second-order approximation as shown in Eq.~(\ref{the_degree_of_precision_theorem1}), but the degree of accuracy rapidly improves as the amplitude of the strong fields increases.
Our result also makes it possible to control the interaction parameter by the local fields, and hence it may be also applicable to accurate control of the interaction.

Finally, we showed the cases where the coupling strength is non-zero.
We first discussed the two cases in which the external fields on the mediator system can enhance the generation of the long-distance entanglement; we introduced the \textit{XY} spin chain with the uniform fields and the random fields with the Hamiltonian~(\ref{XYmodel_the long-distance entanglement2}). 
The long-distance entanglement is enhanced by the random fields on average; there are a few cases in which the entanglement is enhanced to be nearly equal to unity.
This means that we can improve the capability of the mediator system to a great extent.
The long-distance entanglement is also highly enhanced near the point $h_0=2\gamma$, which corresponds to the point of the quantum phase transition in the limit $N\to\infty$.
Second, we discuss a case in which the effective fields of the first order remain to be left; namely, the \textit{XX} spin chain with random interactions.
In this system, the entanglement always vanishes in the limit of the weak coupling because the condition~(\ref{Condition_effective_field_zero}) is not satisfied.
A too strong coupling causes the decrease of the purity because of the entanglement with the mediator system, while a too weak coupling increases the relative amplitude of the effective fields to the effective interaction.
The optimal coupling strength is determined so as to make the both effects minimum.
If we choose the coupling strength properly, the long-distance entanglement still remains to some extent.

In conclusion, we have generally researched necessary conditions for the generation of the entanglement.
Our results show that the conditions for the generation are not so strict in the ground states.
For the practical application, however, we cannot consider the low-temperature limit because it is not realized experimentally, and hence the amplitude of the first excitation energy is important.
We should keep a certain coupling strength in order to increase the first excitation energy as much as possible; in the limit of weak coupling, the first excitation energy always vanishes.
For this purpose, it is essential to find or construct a mediator system best suited to the generation of the long-distance entanglement.
We present the possibilities that the capability of the mediator system can be improved by the external manipulation.
However, we do not understand the principle on how to improve the capability of the mediator system.
Then, the analysis of the capability of the mediator system will be the next problem for the practical application of the long-distance entanglement.
There are several trials~\cite{LDE4,modular} aimed at finding efficient mediator systems for the entanglement generation.


\section*{ACKNOWLEDGMENT}
The present author is grateful to Professor Naomichi Hatano for helpful discussions and comments.

\appendix

\section{Relation between Theorem~1 and the quantum discord} \label{appendixA_chapter3}
In Theorem~1, we show a necessary condition for the entanglement to be generated via a mediator system.
In the present section, we answer the following question: if the condition~(\ref{Sep_cond}) is satisfied, can the quantum discord still exist?
The answer to this question is yes and we show an example in the following.

First, we review the definition of the quantum discord~\cite{discord1}.
Let us consider a two spin system with the density matrix $\rho_{1N}$.
The quantum discord $\mathcal{Q}(\rho_{1N})$ between the spins~1 and $N$ is defined as follows:
\begin{eqnarray}
\mathcal{Q}(\rho_{1N}) \equiv \mathcal{I}(\rho_{1N})  -\mathcal{J}(\rho_{1N})  , \label{Quantum_discord_definition}
\end{eqnarray}
where $\mathcal{I(\rho)}$ is the quantum mutual information defined by
\begin{eqnarray}
\mathcal{I}(\rho) \equiv S(\rho_1) + S(\rho_N) - S(\rho_{1N})   
\end{eqnarray}
with $S(\rho)$ the von Neumann entropy $S(\rho)\equiv\tr (\rho \ln \rho)$.
On the other hand, $\mathcal{J}(\rho)$ is the optimized classical mutual information, which is the maximum information obtained from the measurement of the spins~1 or $N$, and is defined by
\begin{eqnarray}
\mathcal{J}(\rho) \equiv S(\rho_N) -\min_{\Pi_j} \sum_j p_j S(\rho_{N|\Pi_j})   ,
\end{eqnarray}
where $S(\rho_N)$ is the initial von Neumann entropy of the spin~$N$ and  $\sum_j p_j S(\rho_{N|\Pi_j})$ is the average of the von Neumann entropy after the measurement of the spin~1 in the basis of $\Pi_j$.
If the quantum discord~(\ref{Quantum_discord_definition}) has a non-zero value, the correlation between these two spins may not be explained by classical theory.

Let us consider the Hamiltonian
\begin{eqnarray}
H_{{\rm tot}}&=H_{{\rm int}}  + H_{\rm LF}  ,
\end{eqnarray}
where
\begin{eqnarray}
H_{{\rm int}}&=\sigma_1^x   \sigma_{2}^x + \sigma_2^x   \sigma_{3}^x , \nonumber \\
H_{\rm LF} &=\sum_{i=x,y,z}\bigl( h_1^i \sigma_1^i  + h_3^i \sigma_3^i \bigr) .\label{Ising_Hamiltonian_appendix}
\end{eqnarray}
The Hamiltonian~(\ref{Ising_Hamiltonian_appendix}) satisfies the condition~(\ref{Sep_cond}) as
\begin{eqnarray}
&[H_A(\sigma_1), H_B(\sigma_3)]=0, \nonumber \\
&H_A=\sigma_1^x   \sigma_{2}^x , \quad  H_B=\sigma_2^x   \sigma_{3}^x   .
\end{eqnarray}
Therefore, the entanglement can never exist between the spins~1 and 3 in the ground state of $H_{\tot}$ for any values of the local fields $\{h_1^i,h_3^i\}_{i=x,y,z}$.
Indeed,  the density matrix $\rho_{13}$ is given by
\begin{eqnarray}
\rho_{13}=\lim_{\beta\to\infty}\tr_{13}\biggl(\frac{e^{-\beta H_{{\rm tot}}}}{Z_{\tot}(\beta)} \biggr) =
\left( 
\begin{array}{cccc}
0.7286&0&0&0.1250 \\
0&0.1250&0.1250&0\\
0&0.1250&0.1250&0\\
0.1250&0&0&0.02145         
\end{array} 
\right),                      \label{matrix_zero_discord}   
\end{eqnarray}
for $h_1^z=h_3^z=1$ and $h_1^x=h_3^x=h_1^y=h_3^y=0$.
This system has no entanglement.

However, it has a non-zero quantum discord.
We utilize the criterion in Ref.~\cite{discord2} to prove this.
First, we separate the density matrix into the following four blocks:
\begin{eqnarray}
&\rho^{11}= 
\left( 
\begin{array}{cc}
0.7286&0 \\
0&0.1250\\                    
\end{array} 
\right)     ,\ 
\rho^{12}=  \left( 
\begin{array}{cc}
0&0.1250 \\
0.1250&0\\                    
\end{array} 
\right) , \nonumber \\
&\rho^{21}=  \left( 
\begin{array}{cc}
0&0.1250 \\
0.1250&0\\                      
\end{array} 
\right)  ,\ 
\rho^{22}= \left( 
\begin{array}{cc}
0.1250&0 \\
0&0.02145  \\                    
\end{array} 
\right)   .
\end{eqnarray}
A necessary and sufficient condition for the zero discord is given by the following two statements:
\begin{eqnarray}
[\rho^{ij}, (\rho^{ij})^\dagger] =0 \ {\rm for} \ i,j=1,2 \label{Criterion_1_for_quantum_discord}
\end{eqnarray}
and 
\begin{eqnarray}
[\rho^{ij}, \rho^{i'j'}] =0 \ {\rm for} \ i,j,i',j'=1,2. \label{Criterion_2_for_quantum_discord}
\end{eqnarray}
The density matrix (\ref{matrix_zero_discord}) satisfies the first condition (\ref{Criterion_1_for_quantum_discord}) because it is a real matrix.
However, the second (\ref{Criterion_2_for_quantum_discord}) condition is not satisfied. Indeed,
\begin{eqnarray}
\rho^{11}\rho^{12} =\left( 
\begin{array}{cc}
0&0.09107 \\
0.01562&0\\                    
\end{array} 
\right) ,\  \rho^{12}\rho^{11} =\left( 
\begin{array}{cc}
0&0.01562 \\
0.09107&0\\                    
\end{array} 
\right),
\end{eqnarray}
and we have $\rho^{11}\rho^{12}\neq \rho^{12}\rho^{11}$.
Therefore, there exists a quantum discord between the spins~1 and 3.
This shows that the condition in Theorem~1 is applicable only to the existence of the entanglement, not the quantum discord.
So far, we are not sure whether there exists a condition for the indirect interaction to generate a quantum discord.

%



\section*{References}
\renewcommand{\refname}{\vspace{-1cm}}

\end{document}